\documentclass[10pt,american,3p]{elsarticle}
\usepackage[T1]{fontenc}
\usepackage[latin9]{inputenc}
\usepackage{color}
\usepackage{textcomp}
\usepackage{amsmath}
\usepackage{amssymb}
\usepackage{mathdots}
\usepackage{graphicx}
\usepackage{esint}
\PassOptionsToPackage{version=3}{mhchem}
\usepackage{mhchem}

\makeatletter

\providecommand{\tabularnewline}{\\}

\usepackage{pslatex}
\usepackage{graphicx}
\usepackage{amsbsy}
\journal{Computers And Mathematics With Applications}
\usepackage{fontenc}
\usepackage{mathrsfs}
\renewcommand{\boldsymbol}[1]{\pmb{#1}} 
\biboptions{sort&compress}

\AtBeginDocument{
  
}

\makeatother

\usepackage{babel}
\begin{document}
\begin{frontmatter}

\title{Lattice Boltzmann simulations of 3D crystal growth: Numerical schemes for a phase-field model with anti-trapping current}

\author[CEA]{Alain Cartalade\corref{cor1}}

\ead{alain.cartalade@cea.fr}

\author[CEA]{Amina Younsi}

\ead{amina.younsi@cea.fr}

\author[CNRS]{Mathis Plapp}

\ead{Mathis.Plapp@Polytechnique.fr}

\cortext[cor1]{Corresponding author. Tel.:+33 (0)1 69 08 40 67}

\address[CEA]{CEA--Saclay, DEN, DM2S, STMF, LMSF, F-91191 Gif-sur-Yvette, France}

\address[CNRS]{Laboratoire PMC -- Ecole Polytechnique, F-91128 Palaiseau, France}

\begin{abstract}

A lattice-Boltzmann (LB) scheme, based on the Bhatnagar-Gross-Krook
(BGK) collision rules is developed for a phase-field model of alloy
solidification in order to simulate the growth of dendrites. The solidification
of a binary alloy is considered, taking into account diffusive transport
of heat and solute, as well as the anisotropy of the solid-liquid
interfacial free energy. The anisotropic terms in the phase-field
evolution equation, the phenomenological anti-trapping current (introduced
in the solute evolution equation to avoid spurious solute trapping),
and the variation of the solute diffusion coefficient between phases,
make it necessary to modify the equilibrium distribution functions
of the LB scheme with respect to the one used in the standard method
for the solution of advection-diffusion equations. The effects of
grid anisotropy are removed by using the lattices D3Q15 and D3Q19
instead of D3Q7. The method is validated by direct comparison of the
simulation results with a numerical code that uses the finite-difference
method. Simulations are also carried out for two different anisotropy
functions in order to demonstrate the capability of the method to
generate various crystal shapes.

\end{abstract}


\end{frontmatter}

\section{\label{sec:Introduction}Introduction}

With its local collision rules and its easy numerical implementation,
the Lattice Boltzmann Equation (LBE) \citep{Chen-Doolen_AnnRevFlMech1998,Guo-Shu_LBMBook_2013}
is a very attractive method to simulate the dynamics of complex fluids.
Indeed, over more than twenty years, the LBE was successfully applied
to simulate various problems of fluid dynamics, including two-phase
flows separated by an interface \citep{Lee_Parasitic_CAMWA2009,Lee-Liu_DropImpact_LBM_JCP2010}.
Other applications, such as flow and transport in unsaturated porous
media \citep{Ginzburg_AdWR2005a,Ginzburg_PRE2008,Genty-Pot_TRT_TiPM2013},
hydrodynamics coupled with magnetism \citep{Dellar_JCP2002,Pattison_etal_FusEngDes2008},
and even solidification processes \citep{Jiaung_etal_NHT-B_2001}
were also developed.

The phase-field method has become, in recent years, one of the most
popular methods for simulations of crystal growth and microstructure
evolution in materials \citep{Boettinger_etal_AnnRevMatRes2002,SingerLoginova_RepProgPhys2008,Provatas-Elder_Book_2010}.
In this approach, the geometry of domains and interfaces is described
by one or several scalar functions, the phase fields, that take constant
values within each domain and vary smoothly but rapidly through the
interfaces. The evolution equations for the phase fields, which give
the interface dynamics without the need for an explicit front-tracking
algorithm, are nonlinear partial differential equations (PDEs) that
can be obtained from the principles of out-of-equilibrium thermodynamics.
Therefore, they also naturally incorporate thermodynamic boundary
conditions at the interfaces, such as the Gibbs-Thomson condition.
Moreover, it is straightforward to introduce interfacial anisotropy
in phase-field models, which makes it possible to perform accurate
simulations of dendritic growth.

In problems of crystal growth, fluid flow often plays a dominant role.
Indeed, the transport of heat and components from or to the growing
crystal creates density variations in the liquid that trigger natural
convection. Fluid flow may also be induced by external fields (temperature
gradients, magnetic stirring etc.). Therefore, a complete description
of crystal growth requires the coupling of the growth model with a
fluid flow solver. Several phase-field models for solidification that
are coupled to the Navier-Stokes equations for fluid flow have been
proposed in the literature (\citep{Beckermann_etal_JCP1999,Anderson_etal_Convection_PhysD2000,Conti_PRE2001}).
In most cases, direct numerical simulations of these equations with
finite-difference or finite-element methods are employed to solve
the coupled model (see for example \citep{Tonhardt-Amberg_PRE2000,Tong_etal_PRE2001,Jeong_etal_3DsimConv_PRE2001,Lu-Beckermann-Ramirez_JCG2005}).

Many works also exist in the literature that combine the lattice Boltzmann
method with models of solidification or crystal growth \citep{Medvedev-Kassner_LBMCrystGrowthFlows_PRE2005,Rasin-Miller-Succi_PhaseField-CrystGrowth_PRE2005,Chatterjee-Chakraborty_PhysLettA2006,Miller_etal_BinaryAlloy_PhysA2006,Medvedev_etal_PRE2006,Huber_etal_IJHFF2008,Sun_etal_ActaMater2009,Lin_CrystalMagnetism_CF2014}.
Nevertheless, in those papers, the LBE is often used to simulate the
fluid flow only, whereas the model of phase change is simulated with
another numerical method (e.g. finite difference). In some examples
where the LBE is applied to simulate solidification in the presence
of interfacial anisotropy, the model used to track the interface between
the solid and the liquid is not based on the phase-field theory. For
instance in \citep{Sun_etal_ActaMater2009}, the Gibbs-Thomson condition
at the interface is explicitly solved in the numerical procedure,
which corresponds to a <<sharp interface>> method. In \citep{Jiaung_etal_NHT-B_2001,Huber_etal_CoupledDiffusion_JCP2010,Chatterjee-Chakraborty_PhysLettA2006}
the model is based on the <<enthalpy-porosity>> approach, an alternative
model of solid/liquid phase transition for a pure substance \citep{Voller_etal_IJNME1987,Brent_etal_NHT1988}.

Here, we propose a lattice Boltzmann scheme for a phase-field model
of binary alloy solidification \citep{Ramirez_etal_BinaryAlloy_PRE2004}
that takes into account diffusive transport of heat and solute, as
well as the anisotropy of the solid-liquid interfacial free energy.
For this phase-field model, the relationships with its equivalent
<<sharp interface>> equations are well established. When the diffusion
coefficient is not the same in the solid and the liquid, the corrections
of the <<thin interface limit>> of the phase-field model require
adding a phenomenological flux, the anti-trapping current \citep{Karma_PRL2001,Echebarria_etal_PRE2004}.
This model is chosen as a reference by many authors, or used as a
basis by others (see \citep{Plapp_DirectionalSolidification_JCG2007,Provatas-Elder_Book_2010}
for a pedagogical presentation and \citep{Ohno-Matsuura_PRE2009,Galenko_etal_SoluteTrapping_PRE2011,Ohno_PRE2012}
for extensions).

For the development of our scheme, we start from existing LBE formulations
for reaction-diffusion equations. Those are based on the same steps
as the LBE for fluid flow: streaming and collision. In order to apply
this formalism to the equations of the phase-field model, several
modifications must be made. In particular, the presence of \emph{i)}
interfacial anisotropy and \emph{ii)} the anti-trapping current require
to choose appropriate equilibrium distribution functions, to be used
in the collision step. The choice of these functions is dictated by
analytical calculations (a Chapman-Enskog expansion of the LBE equation).

We perform various tests to validate our new scheme. First, we check
the influence of the grid anisotropy on simulated crystal shapes.
We find that for lattices with a sufficient number of streaming directions,
this anisotropy is very low (a fraction of a percent). Next, we directly
compare simulations of dendritic growth in a pure substance and in
an isothermal binary alloy to respective simulations performed with
a finite-difference scheme used in the literature \citep{Karma-Rappel_PRE1998,Karma_PRL2001}.
We find excellent agreement. Finally, we also demonstrate that, in
agreement with previous studies \citep{Karma_Orientation_Nature2006},
our scheme can produce various dendritic shapes (with different growth
directions of the main branches) if the anisotropy function is changed.

As a result, we achieve a full implementation of the phase-field model
in the LBE framework. This has some interesting properties, such as
easy implementation and straightforward parallelization. In addition,
the same concepts involved in fluid dynamics (definitions of lattices,
collision, displacement, bounce-back ...) can be applied, such that
a seamless and easy integration with a LBE solver for fluid flow becomes
possible.

The rest of this paper is organized as follows. The phase-field model
for solidification of a dilute binary mixture is presented in Section
\ref{sec:Mathematical-Model}. The lattice Boltzmann scheme and details
about the algorithm implementation are described in Section \ref{sec:Lattice-Boltzmann-method}.
Section \ref{sec:Validations} presents results of validations and
simulations. Finally, the conclusions are presented in Section \ref{sec:Conclusion}.

\section{\label{sec:Mathematical-Model}Phase-field model}

We consider a phase-field model for the solidification of a single
crystal from a quiescent melt; the fluid is considered at rest and
the density is assumed to be a constant, equal in the liquid and the
solid. Fluid flow is not taken into account in the model. The details
of the model development can be found in \citep{Ramirez_etal_BinaryAlloy_PRE2004};
here, we will only summarize the most important points. The sharp-interface
problem, formulated in terms of the local alloy composition $c$ and
temperature $T$ is:

\begin{subequations}

\begin{align}
\partial_{t}c & =D\boldsymbol{\nabla}^{2}c & \mbox{{\rm (liquid)}},\qquad\\
\partial_{t}T & =\kappa\boldsymbol{\nabla}^{2}T & \mbox{{\rm (liquid and solid)}},\qquad\\
c_{l}(1-k)V_{n} & =-D\partial_{n}c_{l} & \mbox{{\rm (interface)}},\qquad\\
LV_{n} & =C_{p}\kappa(\partial_{n}T|_{s}-\partial_{n}T|_{l}) & \mbox{{\rm (interface)}},\qquad\\
T_{i} & =T_{m}+mc_{l}-\Gamma{\cal K}-V_{n}/\mu_{k} & \mbox{{\rm (interface)}}.\qquad
\end{align}

\end{subequations}

The first two of these equations describe diffusive transport of heat
and solute according to Fick's and Fourier's laws, with $D$ the solute
diffusion coefficient, and $\kappa$ the thermal diffusivity. The
latter, as well as the specific heat $C_{p}$, is assumed to be the
same in the two phases (symmetric model). In contrast, solute transport
is assumed to take place in the liquid only (one-sided model). The
next two equations express mass and heat conservation at the moving
boundary (Stefan conditions), with $V_{n}$ the normal velocity of
the interface, $k=c_{s}/c_{l}$ the partition coefficient that relates
the compositions of solid and liquid in contact with each other at
the interface, $L$ the latent heat of melting, and the symbol $\partial_{n}$
denoting the spatial derivative in the direction normal to the interface.
Indeed, in the phase diagram for a dilute binary mixture, the crystal
has a lower solute concentration than the liquid, so that solute has
to be redistributed upon interface motion. The latent heat of melting
is also set free upon crystallization and generates heat fluxes away
from the interface. The last equation is the Gibbs-Thomson boundary
condition, which relates the interface temperature to the composition
of the adjacent liquid $c_{l}$, the interface curvature ${\cal K}$
and the interface velocity. Here, $T_{m}$ is the melting temperature
of the pure solvent, $m$ the slope of the liquidus line in the phase
diagram, $\Gamma=\gamma T_{m}/L$ the Gibbs-Thomson constant, with
$\gamma$ being the solid-liquid surface free energy, and $\mu_{k}$
is the interface mobility. Note that, for simplicity, we have written
down here the isotropic version of the Gibbs-Thomson condition.

For the following, it is useful to introduce scaled fields: 
\begin{align}
\theta & =\frac{T-T_{m}-mc_{\infty}}{L/C_{p}},\\
U & =\frac{c-c_{\infty}}{(1-k)c_{\infty}},\label{Usharp}
\end{align}
where $c_{\infty}$ is the initial composition of the melt. In terms
of these fields, the equations become:

\begin{subequations}

\begin{align}
\partial_{t}U & =D\boldsymbol{\nabla}^{2}U,\\
\partial_{t}\theta & =\kappa\boldsymbol{\nabla}^{2}\theta,\\
{}[1+(1-k)U_{i}]V_{n} & =-D\partial_{n}U,\\
V_{n} & =\kappa(\partial_{n}\theta|_{s}-\partial_{n}\theta|_{l}),\\
\theta_{i}+Mc_{\infty}U_{i} & =-d_{0}{\cal K}-\beta V_{n}.
\end{align}

\end{subequations}

Here, quantities evaluated at the interface have a subscript $i$,
$M=-m(1-k)C_{p}/L$ is the scaled magnitude of the liquidus slope,
\begin{equation}
d_{0}=\frac{\Gamma C_{p}}{L}=\frac{\gamma T_{m}C_{p}}{L^{2}}
\end{equation}
is the capillary length, with $\gamma$ the solid-liquid surface energy,
and 
\begin{equation}
\beta=\frac{C_{p}}{L\mu_{k}}
\end{equation}
the interface kinetic coefficient.

In the phase-field formulation of this problem \citep{Ramirez_etal_BinaryAlloy_PRE2004},
the interface position is implicitly described as a level set of a
phase-field function $\phi$. The phase field takes the value $\phi=1$
in the solid and $\phi=-1$ in the liquid. Furthermore, the field
$U$ is expressed in terms of $\phi$ and $c$ as 
\begin{equation}
U=\frac{\frac{c/c_{\infty}}{\frac{1}{2}[1+k-(1-k)\phi]}-1}{1-k}.
\end{equation}
This definition extends $U$ to the entire domain (solid, liquid,
and interfaces); in the liquid, it is identical to Eq. (\ref{Usharp}).
At equilibrium, $U$ is constant across the diffuse interface. In
fact, $U$ is a scaled diffusion potential (see \citep{Plapp_PRE2011}
for details).

The model consists of three coupled partial differential equations
for the three fields $\phi$, $\theta$, and $U$ which read:

\begin{subequations}

\begin{align}
\tau(\mathbf{n})\frac{\partial\phi}{\partial t} & =W_{0}^{2}\boldsymbol{\nabla}\cdot(a_{s}^{2}(\mathbf{n})\boldsymbol{\nabla}\phi)+W_{0}^{2}\boldsymbol{\nabla}\cdot\boldsymbol{\mathcal{N}}+(\phi-\phi^{3})-\lambda\left(Mc_{\infty}U+\theta\right)(1-\phi^{2})^{2},\label{eq:PhaseField_KarmaRappel}\\
\left(\frac{1+k}{2}-\frac{1-k}{2}\phi\right)\frac{\partial U}{\partial t} & =\boldsymbol{\nabla}\cdot\left(Dq(\phi)\boldsymbol{\nabla}U-\mathbf{j}_{\mbox{at}}\right)+\left[1+\left(1-k\right)U\right]\frac{1}{2}\frac{\partial\phi}{\partial t},\label{eq:Concentration_Echebarria}\\
\frac{\partial\theta}{\partial t} & =\kappa\boldsymbol{\nabla}^{2}\theta+\frac{1}{2}\frac{\partial\phi}{\partial t}.\label{eq:Temp_Ramirez}
\end{align}

\end{subequations}

Here, $W_{0}$ denotes the characteristic width of the diffuse interfaces,
and the coefficient $\lambda$ describes the strength of the coupling
between the phase field and the transport fields. The relaxation time
of the phase field is noted $\tau(\mathbf{n})$ and depends on the
unit normal vector at the interface $\mathbf{n}=-\boldsymbol{\nabla}\phi/\bigl|\boldsymbol{\nabla}\phi\bigr|$.
We choose $\tau(\mathbf{n})=\tau_{0}\ensuremath{a_{s}^{2}}(\mathbf{n})$,
where $\tau_{0}$ is a constant and $a_{s}(\mathbf{n})$ is an anisotropy
function. For most of the following, we use the standard choice:

\begin{equation}
a_{s}(\mathbf{n})=1-3\varepsilon_{s}+4\varepsilon_{s}\sum_{\alpha=x,y,z}n_{\alpha}^{4},\label{eq:Anisotropic-Function_As}
\end{equation}
which describes a cubic anisotropy of strength $\varepsilon_{s}$
in three dimensions, with $n_{\alpha}$ ($\alpha=x,\, y,\, z$) being
the Cartesian $\alpha$-component of $\mathbf{n}$. The presence of
the anisotropy on the right-hand side of Eq. (\ref{eq:PhaseField_KarmaRappel})
arises from an anisotropic surface free energy; the function $\boldsymbol{\mathcal{N}}\equiv\boldsymbol{\mathcal{N}}(\mathbf{x},\, t)$
is a vector defined by:

\begin{equation}
\boldsymbol{\mathcal{N}}(\mathbf{x},\, t)=\bigl|\boldsymbol{\nabla}\phi\bigr|^{2}a_{s}(\mathbf{n})\left(\frac{\partial a_{s}(\mathbf{n})}{\partial(\partial_{x}\phi)},\,\frac{\partial a_{s}(\mathbf{n})}{\partial(\partial_{y}\phi)},\,\frac{\partial a_{s}(\mathbf{n})}{\partial(\partial_{z}\phi)}\right)^{T}.\label{eq:TermesAnisotropes}
\end{equation}
Expressions of the derivatives $\partial a_{s}(\mathbf{n})/\partial(\partial_{\alpha}\phi)$
will be specified in subsection \ref{sub:LB_D3Q7-D3Q15-D3Q19}.

In Eq. (\ref{eq:Concentration_Echebarria}), $q(\phi)=(1-\phi)/2$
is a function that interpolates the solute diffusivity between $D$
in the liquid and 0 in the solid. $\mathbf{j}_{\mbox{at}}$ is the
phenomenological anti-trapping current introduced in \citep{Karma_PRL2001}
in order to counterbalance spurious solute trapping without introducing
other thin-interface effects (see also \citep{Almgren_SIAMJourApplMath1999,Echebarria_etal_PRE2004}).
It is defined by:

\begin{equation}
\mathbf{j}_{\mbox{at}}(\mathbf{x},\, t)=-\frac{1}{2\sqrt{2}}W_{0}\left[1+\left(1-k\right)U\right]\times\frac{\partial\phi}{\partial t}\frac{\boldsymbol{\nabla}\phi}{\bigl|\boldsymbol{\nabla}\phi\bigr|}.\label{eq:AntiTrapping_Current}
\end{equation}
This current is proportional to the velocity ($\partial_{t}\phi)$
and the thickness $W_{0}$ of the interface, is normal to the interface,
and pointing from solid to liquid ($-\boldsymbol{\nabla}\phi/\bigl|\boldsymbol{\nabla}\phi\bigr|$).
While the other components of the model can be derived variationally
from an appropriate free-energy functional, the anti-trapping current
was introduced for phenomenological reasons and justified by carrying
out matched asymptotic expansions, which demonstrated that the model
with the anti-trapping current is indeed equivalent to the sharp-interface
problem \citep{Echebarria_etal_PRE2004}. Let us mention that, recently,
an alternative justification for this current has been proposed \citep{Brenner-Boussinot_PRE2012,Fang-Mi_ThermodynamicConsistency_Antitrapping_PRE2013}.
In any case, the matched asymptotic expansions provide a relation
between phase-field and sharp-interface parameters given by:

\begin{subequations}

\begin{align}
d_{0} & =a_{1}\frac{W_{0}}{\lambda},\label{eq:Capillary-Length}\\
\beta & =a_{1}\left(\frac{\tau_{0}}{W_{0}\lambda}-a_{2}\frac{W_{0}}{D}\left[\frac{D}{\kappa}+Mc_{\infty}[1+(1-k)U]\right]\right),\label{eq:Kinetic-Coeff}
\end{align}
with $a_{1}$ and $a_{2}$ being numbers of order unity. For the model
used here, $a_{1}=5\sqrt{2}/8$, and $a_{2}\approx0.6267$. These
relations make it possible to choose phase-field parameters for prescribed
values of the capillary length (surface energy) and the interface
mobility (interface kinetic coefficient). Note that the interface
width $W_{0}$ is a parameter that can be freely chosen in this formulation;
the asymptotic analysis remains valid as long as $W_{0}$ remains
much smaller than any length scale present in the sharp-interface
solution of the considered problem (for example, a dendrite tip radius
in the case of dendritic growth).

\end{subequations}

The model presented above can be seen as a combination of the earlier
phase-field formulations for the symmetric model \citep{Karma-Rappel_PRE1998}
and the one-sided model \citep{Karma_PRL2001,Echebarria_etal_PRE2004},
which have been widely used. More detailed derivations and discussions
of the model equations can be found in these references.

\section{\label{sec:Lattice-Boltzmann-method}Lattice Boltzmann schemes}

Eqs. (\ref{eq:PhaseField_KarmaRappel})--(\ref{eq:Temp_Ramirez})
with the additional relationships (\ref{eq:Anisotropic-Function_As})--(\ref{eq:AntiTrapping_Current})
represent the mathematical model considered in this work. In this
section, the numerical method based on the LBE will be described for
each equation of the model. The LBE is an evolution equation in time
and space of a discrete function, the distribution function of particles,
which is defined over a lattice. The choice of the lattice determines
the number of streaming directions of the distribution function. Once
the LBE is defined, the algorithm can be summarized in three main
operations applied on this distribution function: the first one is
a moving step on the lattice; the second one is a collision step that
relaxes the distribution function towards an equilibrium, the equilibrium
distribution function, with a relaxation rate. Finally, the last stage
is to update the physical variable, such as the dimensionless temperature,
or the phase field, by computing its moment of order zero.

In this section we detail each stage of the method: the LBE will be
presented and the equilibrium distribution functions will be defined
as well as the relaxation rates. Next, various lattices will be introduced
and some details will be given about the algorithm implementation.
For a pedagogical presentation, we start the description with the
LB scheme for the heat equation, because it is the simplest equation
of the model for which the standard LB method can be applied. For
the two other ones, the collision step and the equilibrium distribution
function have to be modified. Derivation of equilibrium distribution
functions, which couples the physical variables and the lattice-dependent
quantities, is the most delicate part of the numerical scheme. The
derivations of such functions necessitate to carry out asymptotic
calculations (Chapman-Enskog expansion) that can be found in \ref{sec:Chapman-Enskog-PhaseField}
and \ref{sec:Appendix_Transport} for Eqs. (\ref{eq:PhaseField_KarmaRappel})
and (\ref{eq:Concentration_Echebarria}) respectively.

\subsection{\label{sub:Heat-equation}Heat equation: standard lattice Boltzmann
scheme}

The heat equation (\ref{eq:Temp_Ramirez}) is a diffusion equation
with a source term. For that equation, the standard LB-BGK equation
is applied:

\begin{subequations}

\begin{equation}
f_{i}(\mathbf{x}+\mathbf{e}_{i}\delta x,\, t+\delta t)=f_{i}(\mathbf{x},\, t)-\frac{1}{\eta_{\theta}}\left[f_{i}(\mathbf{x},\, t)-f_{i}^{(0)}(\mathbf{x},\, t)\right]+w_{i}Q_{\theta}(\mathbf{x},\, t)\delta t,\label{eq:LBE_Eq_Temp}
\end{equation}
where $f_{i}(\mathbf{x},\, t)$ is a distribution function which can
be regarded as an intermediate function introduced to calculate the
dimensionless temperature $\theta$. This latter is calculated by:

\begin{equation}
\theta(\mathbf{x},\, t)=\sum_{i=0}^{N_{pop}}f_{i}(\mathbf{x},\, t),\label{eq:Temperature_LB}
\end{equation}
where the index $i$ identifies the moving directions on a lattice:
$i=0,\,...,\, N_{pop}$ where $N_{pop}$ is the total number of directions.
$\mathbf{e}_{i}$ is the vector of displacement on that lattice and
$w_{i}$ are weights. The quantities $N_{pop}$, $\mathbf{e}_{i}$
and $w_{i}$ are lattice-dependent and will be defined in subsection
\ref{sub:LB_D3Q7-D3Q15-D3Q19}. The time-step is noted $\delta t$
and the space-step is noted $\delta x$ by assuming $\delta x=\delta y=\delta z$.
In Eq. (\ref{eq:LBE_Eq_Temp}), the equilibrium distribution function
$f_{i}^{(0)}$ and the source term $Q_{\theta}$ are given by:

\begin{align}
f_{i}^{(0)}(\mathbf{x},\, t) & =w_{i}\theta(\mathbf{x},\, t),\label{eq:EqFunction_Tempr}\\
Q_{\theta}(\mathbf{x},\, t) & =\frac{1}{2}\frac{\partial\phi}{\partial t}.\label{eq:Source_Tempr}
\end{align}

In such a method, the thermal diffusivity $\kappa$ is related to
the relaxation time of collision $\eta_{\theta}$ by:

\begin{equation}
\kappa=e^{2}\left(\eta_{\theta}-\frac{1}{2}\right)\frac{\delta x^{2}}{\delta t},\label{eq:ThermalDiffusivity}
\end{equation}
where $e^{2}$ is an additional lattice-dependent coefficient which
arises from the second-order moment of $f_{i}^{(0)}$. The values
of $e^{2}$ will be given in subsection \ref{sub:LB_D3Q7-D3Q15-D3Q19}
for several lattices. The index $\theta$ in $Q_{\theta}$ and $\eta_{\theta}$
indicates that both quantities are relative to the heat equation.
In a more general case, the thermal diffusivity $\kappa$ is a function
depending on space and time. In that case, the relationship (\ref{eq:ThermalDiffusivity})
must be inverted and the relaxation parameter has to be updated at
each time step.

The principle of the LB scheme is the following. Once the dimensionless
temperature $\theta$ is known, the equilibrium distribution function
$f_{i}^{(0)}$ is computed by using Eq. (\ref{eq:EqFunction_Tempr}).
The collision stage (right-hand side of Eq. (\ref{eq:LBE_Eq_Temp}))
is next calculated and yields an intermediate distribution function
that will be streamed in each direction (left-hand side of Eq. (\ref{eq:LBE_Eq_Temp})).
Finally after updating the boundary conditions, the new temperature
is calculated by using Eq. (\ref{eq:Temperature_LB}) and the algorithm
is iterated in time. Notice that the scheme is fully explicit: all
terms in the right-hand side of Eq. (\ref{eq:LBE_Eq_Temp}) are defined
at time $t$. Also note that the source term $Q_{\theta}$ involves
the time derivative of the phase field. In practice, the heat equation
must be solved after solving the phase-field equation. At the first
time-step, the derivative can be evaluated thanks to the knowledge
of the phase field and the initial condition. Finally, this scheme
can be easily extended to simulate the Advection-Diffusion Equation
(ADE) by modifying the equilibrium distribution function such as $f_{i}^{(0)\, ADE}=w_{i}\theta\left[1+e^{-2}\mathbf{e}_{i}\cdot\mathbf{v}\delta t/\delta x\right]$
where $\mathbf{v}$ is the advective velocity. Moments of zeroth-,
first- and second-order of $f_{i}^{(0)\, ADE}$ are respectively $\theta$,
$\mathbf{v}\theta\delta t/\delta x$ and $e^{2}\theta\overline{\overline{\mathbf{I}}}$
where $\overline{\overline{\mathbf{I}}}$ is the identity tensor of
rank 2.

\end{subequations}

\subsection{\label{sub:LB_Phase-field-eq}Phase-field equation: modification
of collision stage}

The phase-field equation looks like an ADE with an additional factor
$\tau(\mathbf{n})$ in front of the time derivative. In order to handle
this factor and the divergence term $\boldsymbol{\nabla}\cdot\boldsymbol{\mathcal{N}}$,
the standard LB scheme is modified in the following form:

\begin{subequations}

\begin{equation}
a_{s}^{2}(\mathbf{n})g_{i}(\mathbf{x}+\mathbf{e}_{i}\delta x,\, t+\delta t)=g_{i}(\mathbf{x},\, t)-(1-a_{s}^{2}(\mathbf{n}))g_{i}(\mathbf{x}+\mathbf{e}_{i}\delta x,\, t)-\frac{1}{\eta_{\phi}(\mathbf{x},\, t)}\left[g_{i}(\mathbf{x},\, t)-g_{i}^{(0)}(\mathbf{x},\, t)\right]+w_{i}Q_{\phi}(\mathbf{x},\, t)\frac{\delta t}{\tau_{0}},\label{eq:LBE_Eq_Phase}
\end{equation}

\noindent \begin{flushleft}
with the equilibrium distribution function $g_{i}^{(0)}(\mathbf{x},\, t)$
defined by:
\par\end{flushleft}

\begin{equation}
g_{i}^{(0)}(\mathbf{x},\, t)=w_{i}\left(\phi(\mathbf{x},\, t)-\frac{1}{e^{2}}\mathbf{e}_{i}\cdot\boldsymbol{\mathcal{N}}(\mathbf{x},\, t)\frac{\delta t}{\delta x}\frac{W_{0}^{2}}{\tau_{0}}\right).\label{eq:Feq_Phase}
\end{equation}

In Eq. (\ref{eq:LBE_Eq_Phase}), $g_{i}$ is the distribution function
for the phase field $\phi$ calculated by $\phi=\sum_{i=0}^{N_{pop}}g_{i}$
after the streaming step. Moments of zeroth-, first- and second-order
of the equilibrium distribution function $g_{i}^{(0)}$ are respectively
$\sum_{i=0}^{N_{pop}}g_{i}^{(0)}=\phi$, $\sum_{i=0}^{N_{pop}}g_{i}^{(0)}\mathbf{e}_{i}=-\boldsymbol{\mathcal{N}}\delta tW_{0}^{2}/(\tau_{0}\delta x)$,
and $\sum_{i=0}^{N_{pop}}g_{i}^{(0)}\mathbf{e}_{i}\mathbf{e}_{i}=e^{2}\phi\overline{\overline{\mathbf{I}}}$
where $\overline{\overline{\mathbf{I}}}$ is still the identity tensor
of rank 2 (see \ref{sec:Chapman-Enskog-PhaseField}). The scalar function
$Q_{\phi}(\mathbf{x},\, t)$ is the source term of the phase-field
equation (\ref{eq:PhaseField_KarmaRappel}) defined by:

\begin{align}
Q_{\phi}(\mathbf{x},\, t) & =\left[\phi-\lambda(Mc_{\infty}U+\theta)(1-\phi^{2})\right](1-\phi^{2}).\label{eq:LBE_SourceTerm}
\end{align}

In Eq. (\ref{eq:PhaseField_KarmaRappel}) the coefficient $a_{s}^{2}(\mathbf{n})$
plays a similar role as a <<diffusion>> coefficient depending on
position and time (through $\mathbf{n}$ that depends on $\phi$).
The relaxation time $\eta_{\phi}(\mathbf{x},\, t)$ is a function
of position and time and must be updated at each time step by the
relationship:

\begin{equation}
\eta_{\phi}(\mathbf{x},\, t)=\frac{1}{e^{2}}a_{s}^{2}(\mathbf{n})\frac{W_{0}^{2}}{\tau_{0}}\frac{\delta t}{\delta x^{2}}+\frac{1}{2}.\label{eq:LBE_RelaxationTime}
\end{equation}

\end{subequations}

The lattice Boltzmann scheme for the phase-field equation differs
\textcolor{black}{from} the standard LB method for ADE on two points.
The first difference is the presence in Eq. (\ref{eq:LBE_Eq_Phase})
of \emph{(i)} a factor $a_{s}^{2}(\mathbf{n})$ in front of $g_{i}(\mathbf{x}+\mathbf{e}_{i}\delta x,\, t+\delta t)$
in the left-hand side of Eq. (\ref{eq:LBE_Eq_Phase}) and \emph{(ii)}
an additional term $(1-a_{s}^{2}(\mathbf{n}))g_{i}(\mathbf{x}+\mathbf{e}_{i}\delta x,\, t)$
in the right-hand side. The latter term is non-local in space, i.e.,
it is involved in the collision step at time $t$ and needs the knowledge
of $g_{i}$ at \textcolor{black}{the neighboring} nodes $\mathbf{x}+\mathbf{e}_{i}\delta x$.
Those two terms appear to handle the factor $a_{s}^{2}(\mathbf{n})$
in front of the time derivative $\partial\phi/\partial t$ in Eq.
(\ref{eq:PhaseField_KarmaRappel}). We can see it by carrying out
the Taylor expansions of $g_{i}(\mathbf{x}+\mathbf{e}_{i}\delta x,\, t+\delta t)$
and $g_{i}(\mathbf{x}+\mathbf{e}_{i}\delta x,\, t)$ (see \ref{sec:Chapman-Enskog-PhaseField}).
The method is inspired from \citep{Walsh-Saar_WRR2010}.

The second difference with the LB algorithm for ADE, is the definition
of the equilibrium distribution function $g_{i}^{(0)}$ (Eq. (\ref{eq:Feq_Phase})).
The absence of phase field $\phi(\mathbf{x},\, t)$ in the divergence
term (\ref{eq:PhaseField_KarmaRappel}), explains its presence in
the first term inside the brackets (\ref{eq:Feq_Phase}). Moreover,
note the sign change in front of the scalar product, corresponding
to the sign change of advective term in ADE to $+\boldsymbol{\nabla}\cdot\boldsymbol{\mathcal{N}}$
for the phase-field equation. Finally, the presence of factor $W_{0}^{2}/\tau_{0}$
in Eqs. (\ref{eq:Feq_Phase}) and (\ref{eq:LBE_RelaxationTime}) can
be understood by dividing each term of Eq. (\ref{eq:PhaseField_KarmaRappel})
by $\tau_{0}$ and by comparing this equation with the equation for
moments of $g_{i}^{(0)}$ (see Eq. (\ref{eq:Eq_g0_PhaseField}) in
\ref{sec:Chapman-Enskog-PhaseField}) derived from the asymptotic
expansions of Eq. (\ref{eq:LBE_Eq_Phase}).

\subsection{\label{sub:LB_Transport-eq}Supersaturation equation: modification
of the equilibrium distribution function}

In the usual lattice BGK scheme for ADE, the diffusion coefficient
$Dq(\phi)$ would be related to the relaxation time $\eta_{U}$ with
the relationship $Dq(\phi)=e^{2}(\eta_{U}-1/2)\delta x^{2}/\delta t$.
However, in Eq. (\ref{eq:Concentration_Echebarria}), the interpolation
function $q(\phi)$ cancels the diffusion coefficient inside the solid
part. By following the standard method, the relaxation time would
be equal to $1/2$ in the solid part which would lead to the occurrence
of instabilities of the algorithm. Moreover, another source of instabilities
appeared by applying the non-local method of the previous subsection
for factor $((1+k)-(1-k)\phi)/2\equiv\zeta(\phi)$ in front of the
time derivative. In practice, instabilities of algorithm occurred
for several values of the partition coefficient $k$. In order to
overcome these difficulties, the supersaturation equation was reformulated
in the following way:

\begin{subequations}

\begin{equation}
\frac{\partial U}{\partial t}=\boldsymbol{\nabla}\cdot\left[D\boldsymbol{\nabla}\left(\frac{q(\phi)}{\zeta(\phi)}U(\mathbf{x},\, t)\right)\right]-\boldsymbol{\nabla}\cdot\mathbf{J}_{\mbox{tot}}(\mathbf{x},\, t)+S(\mathbf{x},\, t)+\frac{Q_{U}(\mathbf{x},\, t)}{\zeta(\phi)},\label{eq:Supersaturation_Bis}
\end{equation}
with :

\begin{align}
\mathbf{J}_{\mbox{tot}}(\mathbf{x},\, t) & =D\left[\boldsymbol{\nabla}\left(\frac{q(\phi)}{\zeta(\phi)}\right)+q(\phi)\mathbf{F}(\phi)\right]U+\frac{\mathbf{j}_{\mbox{at}}}{\zeta(\phi)},\label{eq:Def_Jtot}\\
S(\mathbf{x},\, t) & =U\boldsymbol{\nabla}\cdot(Dq(\phi)\mathbf{F}(\phi))+\mathbf{j}_{\mbox{at}}\cdot\mathbf{F}(\phi),\label{eq:Def_S}\\
Q_{U}(\mathbf{x},\, t) & =\left[1+\left(1-k\right)U\right]\frac{1}{2}\frac{\partial\phi}{\partial t},\label{eq:Def_Qu}
\end{align}
where $\mathbf{F}(\phi)=\boldsymbol{\nabla}(1/\zeta(\phi))$. The
relationships (\ref{eq:Supersaturation_Bis})--(\ref{eq:Def_Qu})
arise from successive applications of $\boldsymbol{\nabla}(ab)=a\boldsymbol{\nabla}b+b\boldsymbol{\nabla}a$
and $\boldsymbol{\nabla}\cdot(a\mathbf{c})=a\boldsymbol{\nabla}\cdot\mathbf{c}+\mathbf{c}\cdot\boldsymbol{\nabla}a$
where $a$ and $b$ are two scalar functions and $\mathbf{c}$ is
a vectorial function. Note that the inverse of $\zeta(\phi)$ can
be calculated because this function never vanishes for $k>0$. Indeed
$\zeta(\phi)=k$ if $\phi=+1$, $\zeta(\phi)=1$ if $\phi=-1$ and
varies linearly between those two values for $-1<\phi<+1$.

\end{subequations}

The lattice Boltzmann method for simulating the supersaturation equation
is:

\begin{subequations}

\begin{equation}
h_{i}(\mathbf{x}+\mathbf{e}_{i}\delta x,\, t+\delta t)=h_{i}(\mathbf{x},\, t)-\frac{1}{\eta_{U}}\left[h_{i}(\mathbf{x},\, t)-h_{i}^{(0)}(\mathbf{x},\, t)\right]+w_{i}\left[S(\mathbf{x},\, t)+\frac{Q_{U}(\mathbf{x},\, t)}{\zeta(\phi)}\right]\delta t,\label{eq:LBE_Eq_Transport}
\end{equation}
with an equilibrium distribution function $h_{i}^{(0)}(\mathbf{x},\, t)$
defined as (see \ref{sec:Appendix_Transport}):

\begin{equation}
h_{i}^{(0)}(\mathbf{x},\, t)=A_{i}U(\mathbf{x},\, t)+B_{i}\left(\frac{q(\phi)}{\zeta(\phi)}U(\mathbf{x},\, t)+\frac{1}{e^{2}}\mathbf{e}_{i}\cdot\mathbf{J}_{\mbox{tot}}(\mathbf{x},\, t)\frac{\delta t}{\delta x}\right).\label{eq:Feq_Transport}
\end{equation}

In Eq. (\ref{eq:LBE_Eq_Transport}), $h_{i}$ is the distribution
function for the supersaturation: $U=\sum_{i=0}^{N_{pop}}h_{i}$.
The equilibrium distribution function $h_{i}^{(0)}(\mathbf{x},\, t)$
was derived such as its moments of zeroth-, first- and second-order
are respectively $\sum_{i=0}^{N_{pop}}h_{i}^{(0)}=U$, $\sum_{i=0}^{N_{pop}}h_{i}^{(0)}\mathbf{e}_{i}=\mathbf{J}_{\mbox{tot}}\delta t/\delta x$,
and $\sum_{i=0}^{N_{pop}}h_{i}^{(0)}\mathbf{e}_{i}\mathbf{e}_{i}=e^{2}(q(\phi)/\zeta(\phi))U\overline{\overline{\mathbf{I}}}$
(see \ref{sec:Appendix_Transport}). The values of weights $A_{i}$
and $B_{i}$ are indicated in subsection \ref{sub:LB_D3Q7-D3Q15-D3Q19}
for several lattices. The relaxation time $\eta_{U}$ is calculated
before the time iterations by:

\begin{equation}
\eta_{U}=\frac{1}{e^{2}}\frac{\delta t}{\delta x^{2}}D+\frac{1}{2}.\label{eq:RelaxationTime_Concentration}
\end{equation}

\end{subequations}

With this formulation, the interpolation function $q(\phi)$ and the
relaxation coefficient $\eta_{U}$ are decoupled. Once $\delta x$
and $\delta t$ are fixed, $\eta_{U}$ keeps the same constant value
in the whole computational domain, even in the solid part. The function
$q(\phi)$ appears inside three terms: the laplacian term, the total
flux $\mathbf{J}_{\mbox{tot}}$ and the source term $S$. The second
advantage of this formulation is that the standard collision scheme
can be kept to handle the factor $\zeta(\phi)$ in the LB scheme.
Nevertheless, additional gradients of $\zeta(\phi)$ and $q(\phi)$
have to be evaluated with this formulation.

\subsection{\label{sub:LB_D3Q7-D3Q15-D3Q19}Definitions of lattices and algorithm
implementation}

\paragraph{Definitions of Lattices}

\begin{figure}
\begin{centering}
\begin{tabular}{ccc}
{\footnotesize (a) D3Q7} & {\footnotesize (b) D3Q15} & {\footnotesize (c) D3Q19}\tabularnewline
 &  & \tabularnewline
\includegraphics[scale=0.25]{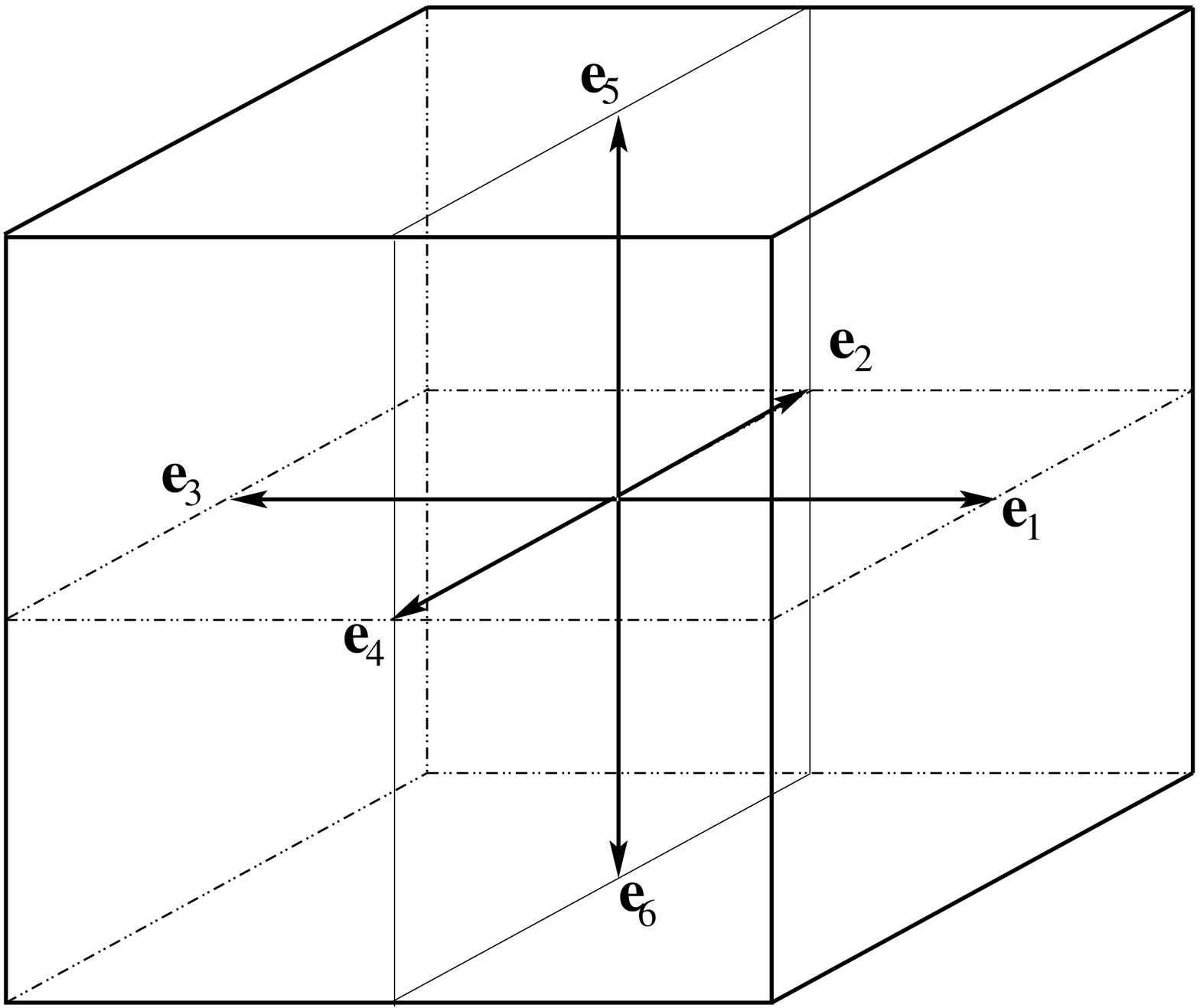} & \includegraphics[scale=0.25]{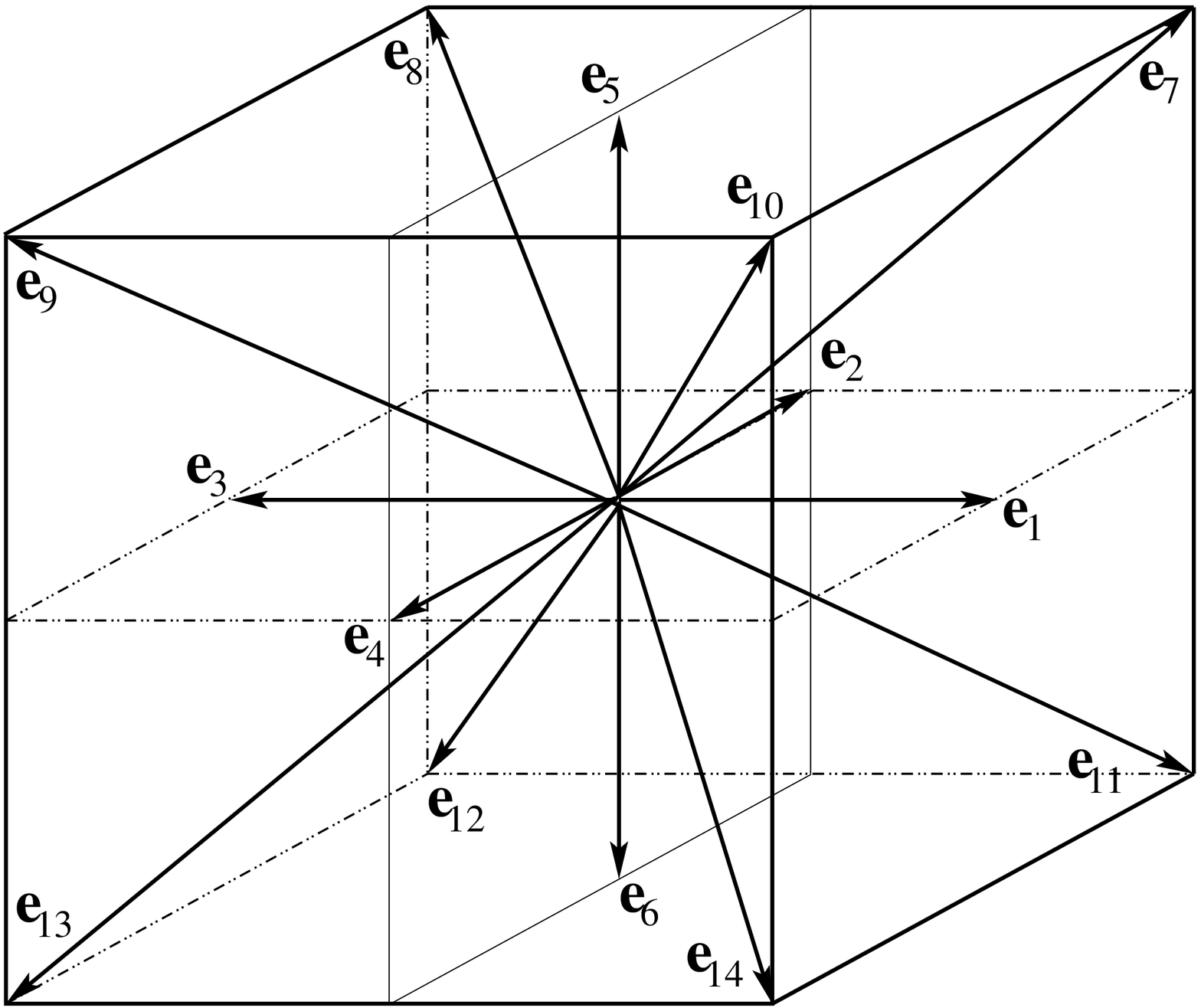} & \includegraphics[scale=0.25]{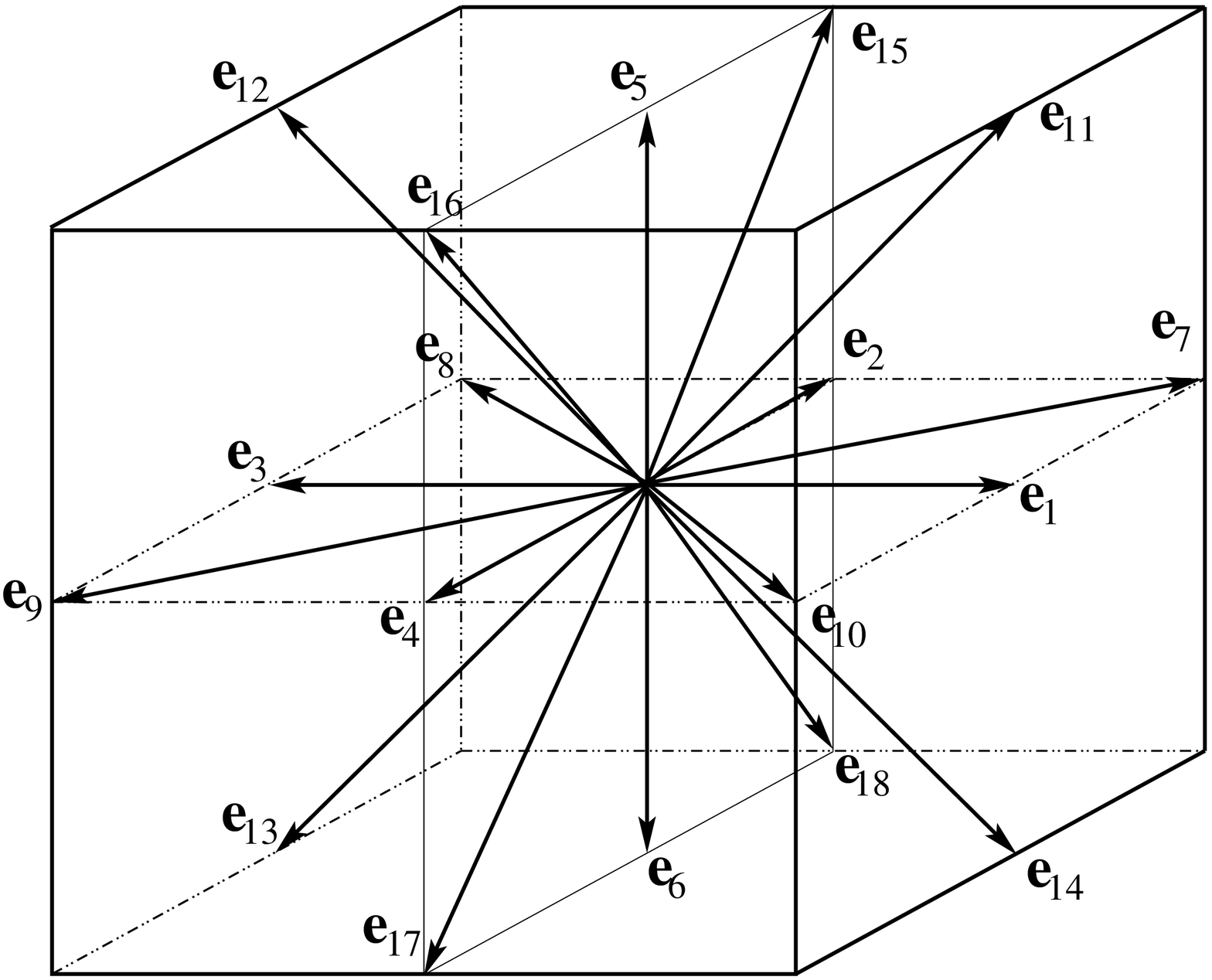}\tabularnewline
\end{tabular}
\par\end{centering}

\caption{\label{fig:Lattice-D3Qb}3D lattices of LB scheme.}
\end{figure}

In order to study the effects of grid anisotropy, which arise from
discretization of the phase-field equation \citep{Karma-Rappel_PRE1998,Bragard_etal_InterfSci2002,Nestler_etal_JCP2005},
three 3D lattices were used in this work: D3Q7, D3Q15 and D3Q19 (Fig.
\ref{fig:Lattice-D3Qb}). The total number of moving directions for
each lattice is respectively $N_{pop}=6,$ $14$ and $18$. The displacement
vectors are defined in Tab. \ref{tab:Def_Vect-E} for all lattices.
The D3Q7-lattice is defined by seven vectors, for D3Q15 eight directions
are added to the previous ones, corresponding to the eight diagonals
of the cube, and for D3Q19 we consider 12 additional directions. For
each one of them, the LB schemes described in the previous subsections
remain identical. The values of weights $w_{i}$, $A_{i}$ , $B_{i}$
and $e^{2}$ are indicated in Tab. \ref{tab:Poids_D3Qb}. For completeness,
we introduce the 2D lattices D2Q5 and D2Q9 for 2D simulations of validation.
The vectors of displacement are defined in Tab. \ref{tab:Def_Vect-E2D}
and the values of weights in Tab. \ref{tab:Poids_D2Qb}.

\begin{table}
\caption{\label{tab:Def_Vect-E}Definition of moving vectors $\mathbf{e}_{i}$
for 3D lattices}

~

\centering{}%
\begin{tabular}{c}
\hline 
{\footnotesize Definition of $\mathbf{e}_{i}$ for D3Q7}\tabularnewline
{\footnotesize }%
\begin{tabular}{ccccccc}
{\footnotesize $\mathbf{e}_{0}=\left(\begin{array}{c}
0\\
0\\
0
\end{array}\right)$} & {\footnotesize $\mathbf{e}_{1}=\left(\begin{array}{c}
1\\
0\\
0
\end{array}\right)$} & {\footnotesize $\mathbf{e}_{2}=\left(\begin{array}{c}
0\\
1\\
0
\end{array}\right)$} & {\footnotesize $\mathbf{e}_{3}=\left(\begin{array}{c}
-1\\
0\\
0
\end{array}\right)$} & {\footnotesize $\mathbf{e}_{4}=\left(\begin{array}{c}
0\\
-1\\
0
\end{array}\right)$} & {\footnotesize $\mathbf{e}_{5}=\left(\begin{array}{c}
0\\
0\\
1
\end{array}\right)$} & {\footnotesize $\mathbf{e}_{6}=\left(\begin{array}{c}
0\\
0\\
-1
\end{array}\right)$}\tabularnewline
\end{tabular}\tabularnewline
\tabularnewline
{\footnotesize Additional $\mathbf{e}_{i}$ vectors for D3Q15}\tabularnewline
{\footnotesize }%
\begin{tabular}{cccc}
{\footnotesize $\mathbf{e}_{7}=\left(\begin{array}{c}
1\\
1\\
1
\end{array}\right)$} & {\footnotesize $\mathbf{e}_{8}=\left(\begin{array}{c}
-1\\
1\\
1
\end{array}\right)$} & {\footnotesize $\mathbf{e}_{9}=\left(\begin{array}{c}
-1\\
-1\\
1
\end{array}\right)$} & {\footnotesize $\mathbf{e}_{10}=\left(\begin{array}{c}
1\\
-1\\
1
\end{array}\right)$}\tabularnewline
{\footnotesize $\mathbf{e}_{11}=\left(\begin{array}{c}
1\\
1\\
-1
\end{array}\right)$} & {\footnotesize $\mathbf{e}_{12}=\left(\begin{array}{c}
-1\\
1\\
-1
\end{array}\right)$} & {\footnotesize $\mathbf{e}_{13}=\left(\begin{array}{c}
-1\\
-1\\
-1
\end{array}\right)$} & {\footnotesize $\mathbf{e}_{14}=\left(\begin{array}{c}
1\\
-1\\
-1
\end{array}\right)$}\tabularnewline
\end{tabular}\tabularnewline
\tabularnewline
{\footnotesize Additional $\mathbf{e}_{i}$ vectors for D3Q19}\tabularnewline
{\footnotesize }%
\begin{tabular}{cccccc}
{\footnotesize $\mathbf{e}_{7}=\left(\begin{array}{c}
1\\
1\\
0
\end{array}\right)$} & {\footnotesize $\mathbf{e}_{8}=\left(\begin{array}{c}
-1\\
1\\
0
\end{array}\right)$} & {\footnotesize $\mathbf{e}_{9}=\left(\begin{array}{c}
1\\
-1\\
0
\end{array}\right)$} & {\footnotesize $\mathbf{e}_{10}=\left(\begin{array}{c}
-1\\
-1\\
0
\end{array}\right)$} & {\footnotesize $\mathbf{e}_{11}=\left(\begin{array}{c}
1\\
0\\
1
\end{array}\right)$} & {\footnotesize $\mathbf{e}_{12}=\left(\begin{array}{c}
-1\\
0\\
1
\end{array}\right)$}\tabularnewline
{\footnotesize $\mathbf{e}_{13}=\left(\begin{array}{c}
1\\
0\\
-1
\end{array}\right)$} & {\footnotesize $\mathbf{e}_{14}=\left(\begin{array}{c}
-1\\
0\\
-1
\end{array}\right)$} & {\footnotesize $\mathbf{e}_{15}=\left(\begin{array}{c}
0\\
1\\
1
\end{array}\right)$} & {\footnotesize $\mathbf{e}_{16}=\left(\begin{array}{c}
0\\
-1\\
1
\end{array}\right)$} & {\footnotesize $\mathbf{e}_{17}=\left(\begin{array}{c}
0\\
1\\
-1
\end{array}\right)$} & {\footnotesize $\mathbf{e}_{18}=\left(\begin{array}{c}
0\\
-1\\
-1
\end{array}\right)$}\tabularnewline
 &  &  &  &  & \tabularnewline
\end{tabular}\tabularnewline
\hline 
\end{tabular}
\end{table}

\begin{table}
\caption{\label{tab:Poids_D3Qb}Values of $w_{i}$, $A_{i}$, $B_{i}$ and
$e^{2}$ for 3D lattices ({\scriptsize $\times$}: irrelevant)}

~

\centering{}%
\begin{tabular}{ccc}
\hline 
\textbf{\scriptsize Lattices} & \textbf{\scriptsize Weights for of $\phi$- and $\theta$-Eq.} & \textbf{\scriptsize Weights for $U$-Eq.}\tabularnewline
\hline 
{\scriptsize }%
\begin{tabular}{ccc}
{\scriptsize Lattice} & {\scriptsize $N_{pop}$} & {\scriptsize $e^{2}$}\tabularnewline
{\scriptsize D3Q7} & {\scriptsize $6$} & {\scriptsize $1/4$}\tabularnewline
{\scriptsize D3Q15} & {\scriptsize $14$} & {\scriptsize $1/3$}\tabularnewline
{\scriptsize D3Q19} & {\scriptsize $18$} & {\scriptsize $1/3$}\tabularnewline
\end{tabular} & {\scriptsize }%
\begin{tabular}{cccc}
{\scriptsize $w_{0}$} & {\scriptsize $w_{1,...,6}$} & {\scriptsize $w_{7,...,14}$} & {\scriptsize $w_{7,...,18}$}\tabularnewline
{\scriptsize $1/4$} & {\scriptsize $1/8$} & {\scriptsize $\times$} & {\scriptsize $\times$}\tabularnewline
{\scriptsize $2/9$} & {\scriptsize $1/9$} & {\scriptsize $1/72$} & {\scriptsize $\times$}\tabularnewline
{\scriptsize $1/3$} & {\scriptsize $1/18$} & {\scriptsize $\times$} & {\scriptsize $1/36$}\tabularnewline
\end{tabular} & {\scriptsize }%
\begin{tabular}{cccccccc}
{\scriptsize $A_{0}$} & {\scriptsize $A_{1,...,6}$} & {\scriptsize $A_{7,...,14}$} & {\scriptsize $A_{7,...,18}$} & {\scriptsize $B_{0}$} & {\scriptsize $B_{1,...,6}$} & {\scriptsize $B_{7,...,14}$} & {\scriptsize $B_{7,...,18}$}\tabularnewline
{\scriptsize $1$} & {\scriptsize $0$} & {\scriptsize $\times$} & {\scriptsize $\times$} & {\scriptsize $-3/4$} & {\scriptsize $1/8$} & {\scriptsize $\times$} & {\scriptsize $\times$}\tabularnewline
{\scriptsize $1$} & {\scriptsize $0$} & {\scriptsize $0$} & {\scriptsize $\times$} & {\scriptsize $-7/9$} & {\scriptsize $1/9$} & {\scriptsize $1/72$} & {\scriptsize $\times$}\tabularnewline
{\scriptsize $1$} & {\scriptsize $0$} & {\scriptsize $\times$} & {\scriptsize $0$} & {\scriptsize $-2/3$} & {\scriptsize $1/18$} & {\scriptsize $\times$} & {\scriptsize $1/36$}\tabularnewline
\end{tabular}\tabularnewline
\hline 
\end{tabular}
\end{table}

\begin{figure}
\begin{centering}
\begin{tabular}{ccc}
{\footnotesize (a) D2Q5} & $\qquad$ & {\footnotesize (b) D2Q9}\tabularnewline
 &  & \tabularnewline
\includegraphics[scale=0.35]{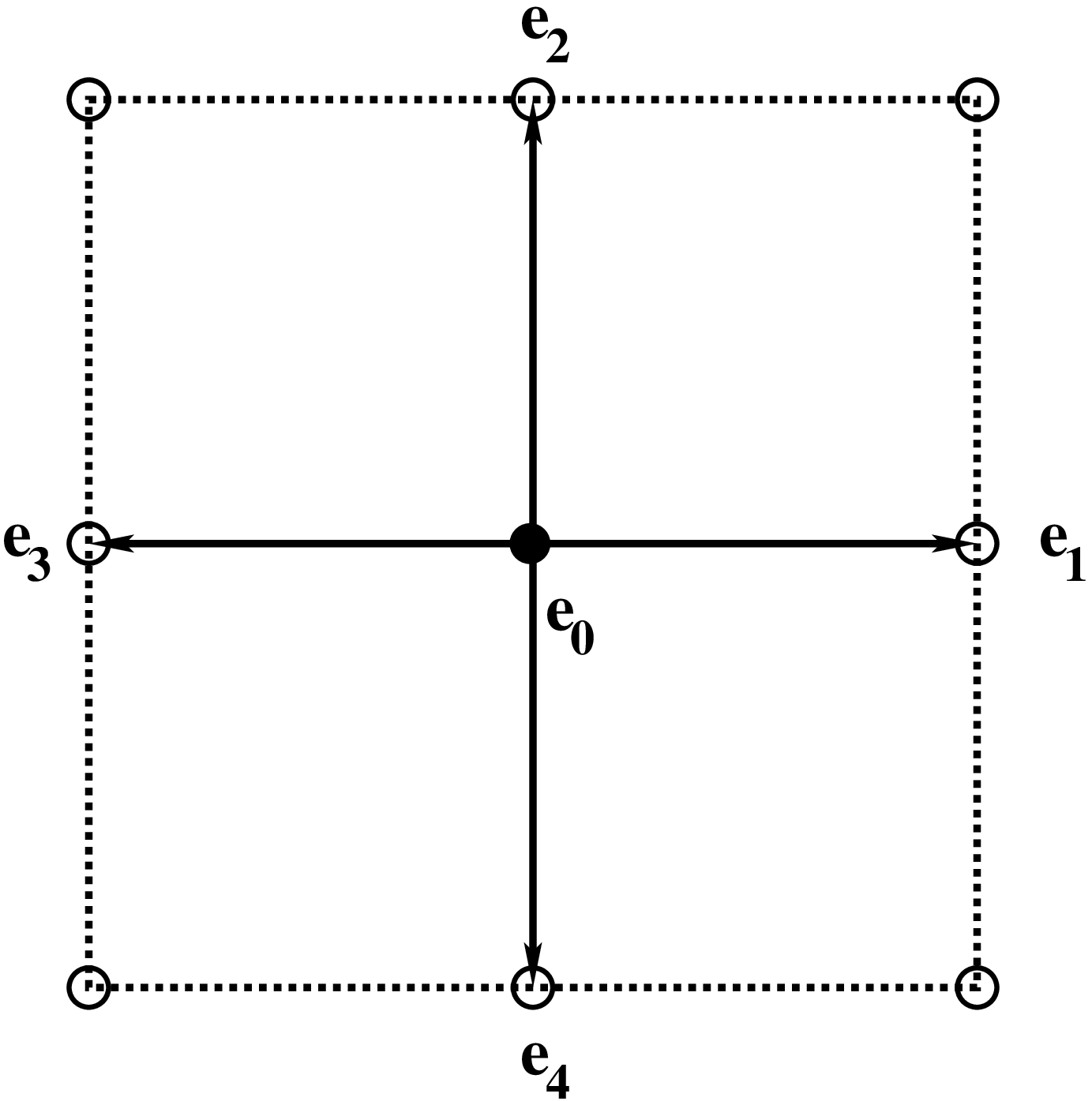} &  & \includegraphics[scale=0.35]{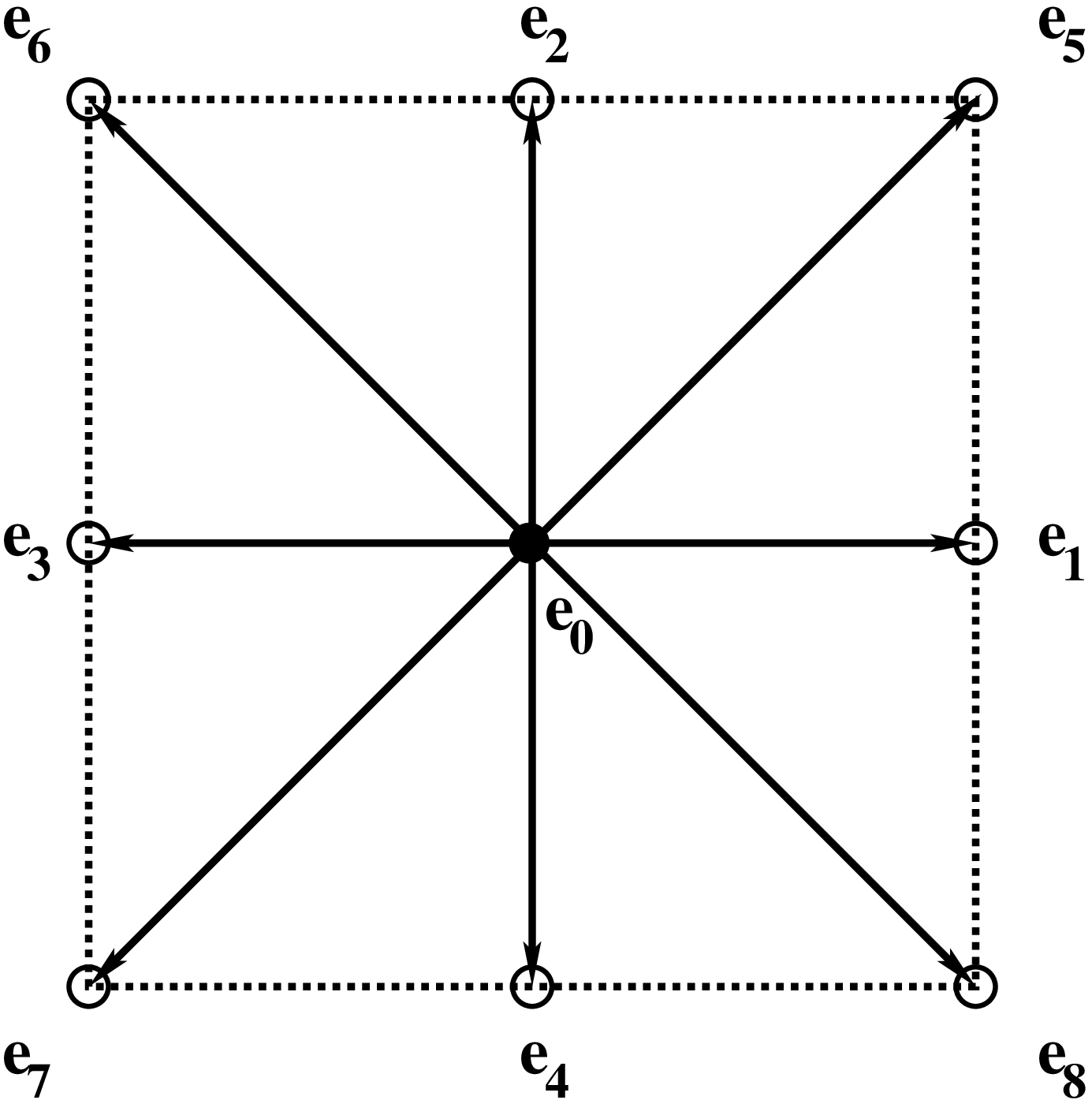}\tabularnewline
\end{tabular}
\par\end{centering}

\caption{\label{fig:Lattice-D2Qb}2D Lattices of LB scheme.}
\end{figure}

\begin{table}
\caption{\label{tab:Def_Vect-E2D}Definition of moving vectors $\mathbf{e}_{i}$
for 2D lattices}

~

\centering{}%
\begin{tabular}{c}
\hline 
{\footnotesize Definition of vectors for D2Q5}\tabularnewline
{\footnotesize }%
\begin{tabular}{ccccc}
{\footnotesize $\mathbf{e}_{0}=\left(\begin{array}{c}
0\\
0
\end{array}\right)$} & {\footnotesize $\mathbf{e}_{1}=\left(\begin{array}{c}
1\\
0
\end{array}\right)$} & {\footnotesize $\mathbf{e}_{2}=\left(\begin{array}{c}
0\\
1
\end{array}\right)$} & {\footnotesize $\mathbf{e}_{3}=\left(\begin{array}{c}
-1\\
0
\end{array}\right)$} & {\footnotesize $\mathbf{e}_{4}=\left(\begin{array}{c}
0\\
-1
\end{array}\right)$}\tabularnewline
\end{tabular}\tabularnewline
\tabularnewline
{\footnotesize Additional vectors for D2Q9}\tabularnewline
{\footnotesize }%
\begin{tabular}{cccc}
{\footnotesize $\mathbf{e}_{5}=\left(\begin{array}{c}
1\\
1
\end{array}\right)$} & {\footnotesize $\mathbf{e}_{6}=\left(\begin{array}{c}
-1\\
1
\end{array}\right)$} & {\footnotesize $\mathbf{e}_{7}=\left(\begin{array}{c}
-1\\
-1
\end{array}\right)$} & {\footnotesize $\mathbf{e}_{8}=\left(\begin{array}{c}
1\\
-1
\end{array}\right)$}\tabularnewline
\end{tabular}\tabularnewline
\tabularnewline
\hline 
\end{tabular}
\end{table}

\begin{table}
\caption{\label{tab:Poids_D2Qb}Values of $w_{i}$, $A_{i}$, $B_{i}$ and
$e^{2}$ for 2D lattices ({\scriptsize $\times$}: irrelevant)}

~

\centering{}{\footnotesize }%
\begin{tabular}{ccc}
\hline 
\textbf{\footnotesize Lattices} & \textbf{\footnotesize Weights for $\phi$- and $\theta$-Eq.} & \textbf{\footnotesize Weights for $U$-Eq.}\tabularnewline
\hline 
{\footnotesize }%
\begin{tabular}{ccc}
{\footnotesize Lattice} & {\footnotesize $N_{pop}$} & {\footnotesize $e^{2}$}\tabularnewline
{\footnotesize D2Q5} & {\footnotesize $4$} & {\footnotesize $1/3$}\tabularnewline
{\footnotesize D2Q9} & {\footnotesize $8$} & {\footnotesize $1/3$}\tabularnewline
\end{tabular} & {\footnotesize }%
\begin{tabular}{ccc}
{\footnotesize $w_{0}$} & {\footnotesize $w_{1,...,4}$} & {\footnotesize $w_{5,...,8}$}\tabularnewline
{\footnotesize $1/3$} & {\footnotesize $1/6$} & {\footnotesize $\times$}\tabularnewline
{\footnotesize $4/9$} & {\footnotesize $1/9$} & {\footnotesize $1/36$}\tabularnewline
\end{tabular} & {\footnotesize }%
\begin{tabular}{cccccc}
{\footnotesize $A_{0}$} & {\footnotesize $A_{1,...,4}$} & {\footnotesize $A_{5,...,8}$} & {\footnotesize $B_{0}$} & {\footnotesize $B_{1,...,4}$} & {\footnotesize $B_{5,...,8}$}\tabularnewline
{\footnotesize $1$} & {\footnotesize $0$} & {\footnotesize $\times$} & {\footnotesize $-2/3$} & {\footnotesize $1/6$} & {\footnotesize $\times$}\tabularnewline
{\footnotesize $1$} & {\footnotesize $0$} & {\footnotesize $0$} & {\footnotesize $-5/9$} & {\footnotesize $1/9$} & {\footnotesize $1/36$}\tabularnewline
\end{tabular}\tabularnewline
\hline 
\end{tabular}
\end{table}

\paragraph{Algorithm implementation}

The algorithm is \textcolor{black}{sequential}: after solving the
phase-field equation, the phase-field $\phi$ is used to calculate
the time evolution of the supersaturation $U$ and the temperature
$\theta$. For each equation, the standard stages of lattice Boltzmann
method are applied. Each LB equation (\ref{eq:LBE_Eq_Temp}), (\ref{eq:LBE_Eq_Phase})
and (\ref{eq:LBE_Eq_Transport}), is \textcolor{black}{separated}
into one collision step followed by one streaming step of each distribution
function $f_{i}$, $g_{i}$ and $h_{i}$. The factors $a_{s}^{2}(\mathbf{n})$
and $\zeta(\phi)$ are treated explicitly. The collision stage for
the phase-field equation writes:

\begin{subequations}

\begin{align}
g_{i}^{\star}(\mathbf{x},\, t) & =\frac{1}{a_{s}^{2}(\mathbf{n})}\biggl\{ g_{i}(\mathbf{x},\, t)-(1-a_{s}^{2}(\mathbf{n}))g_{i}(\mathbf{x}+\mathbf{e}_{i}\delta x,\, t)-\frac{1}{\eta_{\phi}(\mathbf{x},\, t)}\left[g_{i}(\mathbf{x},\, t)-g_{i}^{(0)}(\mathbf{x},\, t)\right]+w_{i}Q(\mathbf{x},\, t)\frac{\delta t}{\tau_{0}}\biggr\},\label{eq:Collision_Stage}
\end{align}
where the symbol $\star$ means the distribution function after the
collision. The standard collision ($a_{s}^{2}(\mathbf{n})=1$) is
considered on boundary nodes. The moving step writes:

\begin{equation}
g_{i}(\mathbf{x}+\mathbf{e}_{i}\delta x,\, t+\delta t)=g_{i}^{\star}(\mathbf{x},\, t).\label{eq:Deplacement_STD}
\end{equation}

\end{subequations}

For each LB scheme, the update of boundary conditions is carried out
by the <<bounce back>> rule. For instance in the phase-field scheme
$g_{i}(\mathbf{x},\, t)=g_{i'}(\mathbf{x},\, t)$ where $i'$ is the
opposite direction of $i$. The computation of gradient $\boldsymbol{\nabla}\phi$
needed for the normal vector $\mathbf{n}$ is carried out by a centered
finite difference method. Finally, the computation of vector $\boldsymbol{\mathcal{N}}(\mathbf{x},\, t)$
for each time step needs the calculation of derivatives $\partial a_{s}(\mathbf{n})/\partial(\partial_{\alpha}\phi)$
for $\alpha=x,\, y,\, z$, which write:

\begin{align}
\frac{\partial a_{s}(\mathbf{n})}{\partial(\partial_{\alpha}\phi)} & =-\frac{16\varepsilon_{s}}{\bigl|\boldsymbol{\nabla}\phi\bigr|^{6}}\times(\partial_{\alpha}\phi)\Bigl[(\partial_{\beta}\phi)^{4}-(\partial_{\alpha}\phi)^{2}(\partial_{\beta}\phi)^{2}-(\partial_{\alpha}\phi)^{2}(\partial_{\gamma}\phi)^{2}+(\partial_{\gamma}\phi)^{4}\Bigr].\label{eq:Derivees_As}
\end{align}

In this equation, the first component of $\boldsymbol{\mathcal{N}}$
is obtained for $\alpha\equiv x$, $\beta\equiv y$ and $\gamma\equiv z$.
The second one is obtained for $\alpha\equiv y$, $\beta\equiv x$
and $\gamma\equiv x$ and finally the third one for $\alpha\equiv z$,
$\beta\equiv x$ and $\gamma\equiv y$. The gradient terms involved
in Eq. (\ref{eq:Def_Jtot}) and (\ref{eq:Def_S}) are calculated with
a centered finite difference method. The partial derivative in time
$\partial\phi/\partial t$ in Eqs. (\ref{eq:Concentration_Echebarria}),
(\ref{eq:Temp_Ramirez}) and (\ref{eq:AntiTrapping_Current}) is discretized
by an Euler scheme.

\section{\label{sec:Validations}Validations and simulations}

For simulations, the computational domain is cubic and zero fluxes
are imposed on all boundaries for each equation. For the phase-field
equation, a nucleus is initialized as a diffuse sphere: $\phi(\mathbf{x},\,0)=\tanh\left[(R_{s}-d_{s})/\sqrt{l_{s}}\right]$
where $R_{s}$ is the radius, $d_{s}=\sqrt{(x-x_{s})^{2}+(y-y_{s})^{2}+(z-z_{s})^{2}}$
and $\mathbf{x}_{s}=(x_{s},\, y_{s},\, z_{s})^{T}$ is the position
of its center. With this initial condition, $\phi=+1$ inside the
sphere and $\phi=-1$ outside. The coefficient $l_{s}$ decreases
or increases the slope of $\phi$-profile between its minimal and
maximal values. In this work $l_{s}=2W_{0}$ as indicated in \citep{Li_etal_OpSplitting_JCG2011}.
For equations of supersaturation and temperature, the initial conditions
are constant on the whole domain: $U(\mathbf{x},\,0)=U_{0}$ and $\theta(\mathbf{x},\,0)=\theta_{0}$.

\subsection{Crystal growth of pure substance: 3D grid effects and validation
with a benchmark}

We consider the basic problem of solidification of a pure substance.
For this problem, the phase-field model is composed of two equations
\citep{Karma-Rappel_PRE1998}, the first one for the phase-field Eq.
(\ref{eq:PhaseField_KarmaRappel}) by setting $M=0$ and the second
one for the dimensionless temperature Eq. (\ref{eq:Temp_Ramirez}).
The lattice Boltzmann schemes of subsections \ref{sub:Heat-equation}
and \ref{sub:LB_Phase-field-eq} are checked with a finite difference
scheme. Following \citep{Bragard_etal_InterfSci2002}, the discrete
laplacian of phase-field equation is obtained by using respectively
6 (FD6) and 18 (FD18) nearest neighboring nodes. Simulations are first
carried out for an isotropic case, i.e. with $\varepsilon_{s}=0$,
for studying the lattice effects. Next, the anisotropic term ($\varepsilon_{s}\neq0$)
will be considered for comparison of the numerical implementation
of LB schemes with another code.

\paragraph{Isotropic case: $\varepsilon_{s}=0$}

\begin{figure*}
\begin{centering}
\begin{tabular}{cc}
{\footnotesize (a) Plane of normal vector $(0,\,0,\,1)^{T}$} & {\footnotesize (b) Plane of normal vector $(1,\,0,\,-1)^{T}$}\tabularnewline
 & {\footnotesize (view in $xy$-plane)}\tabularnewline
\includegraphics[angle=-90,scale=0.33]{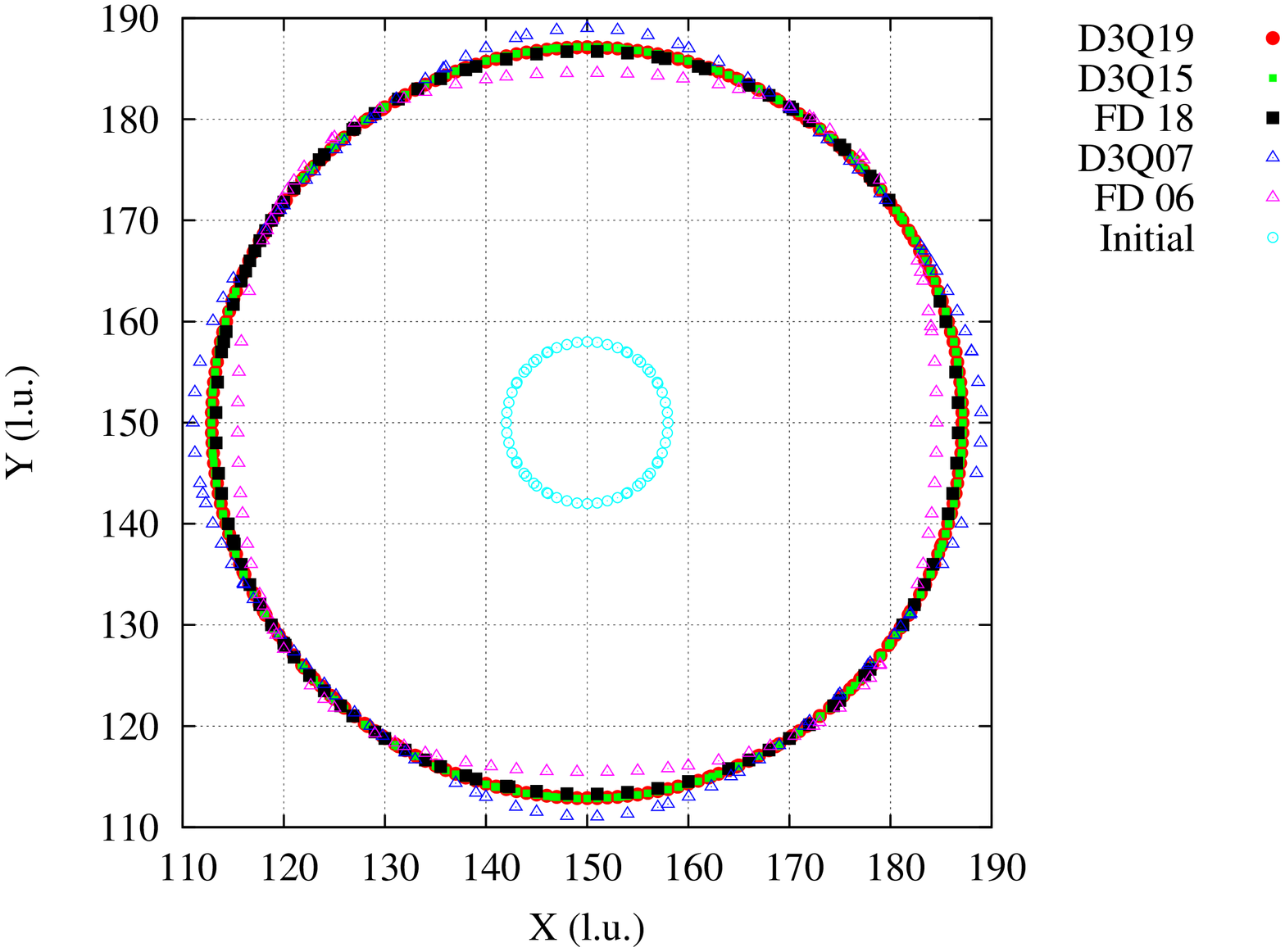} & \includegraphics[angle=-90,scale=0.33]{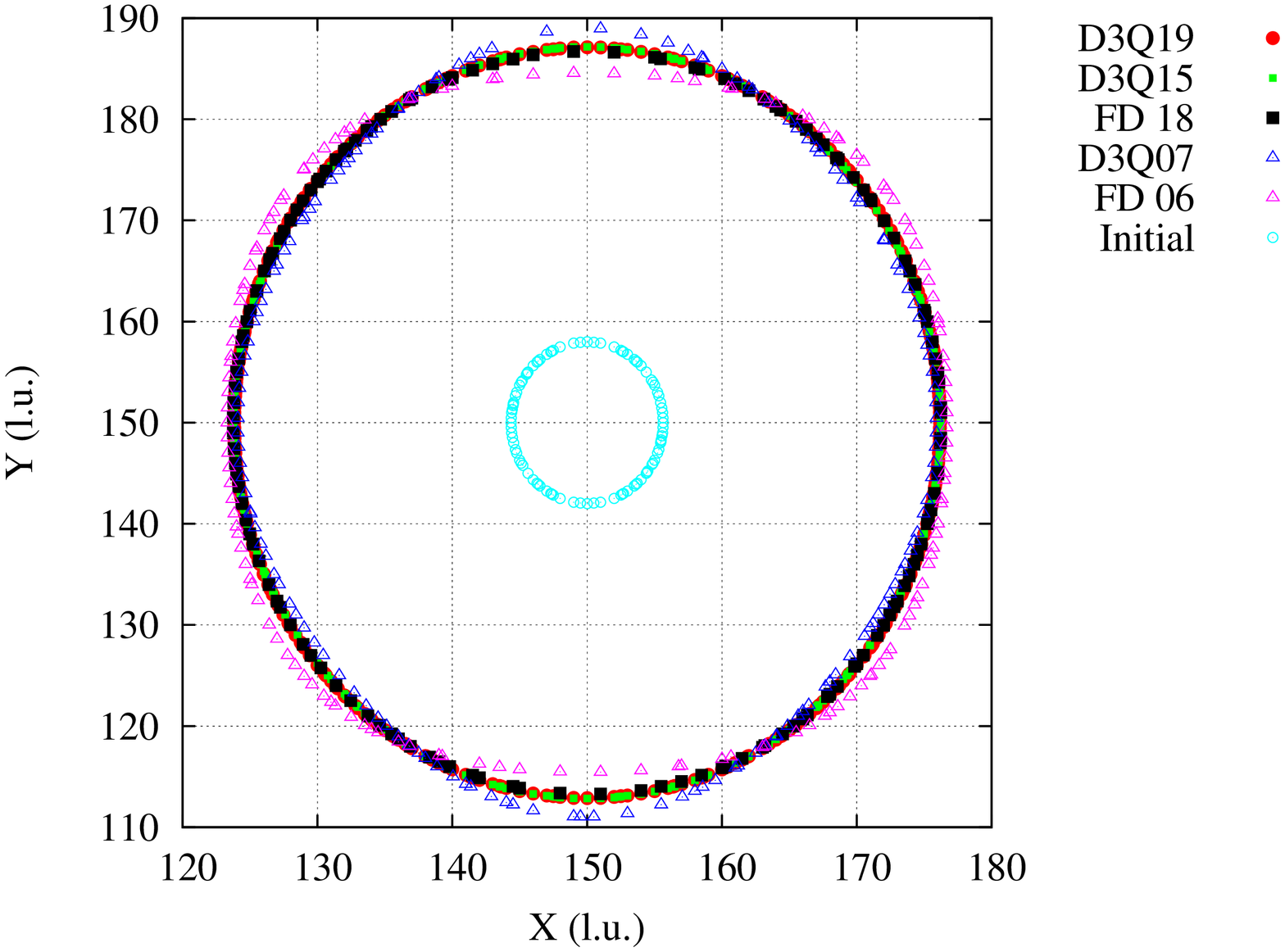}\tabularnewline
\end{tabular}
\par\end{centering}

~

~

\caption{\label{fig:Slices_D3Q7-D3Q19}Iso-values $\phi=0$ of phase field
for LB and FD schemes at $t=10^{4}\delta t$. Results from LB-D3Q15,
LB-D3Q19 and FD18 form circles which match, contrary to those arising
from LB-D3Q7. The initial condition is given for comparison.}
\end{figure*}

\begin{figure*}
\begin{centering}
\begin{tabular}{ccc}
{\footnotesize (a) Direction $(1,\,0,\,0)$} &  & {\footnotesize (b) Direction $(1,\,1,\,0)$}\tabularnewline
\includegraphics[angle=-90,scale=0.33]{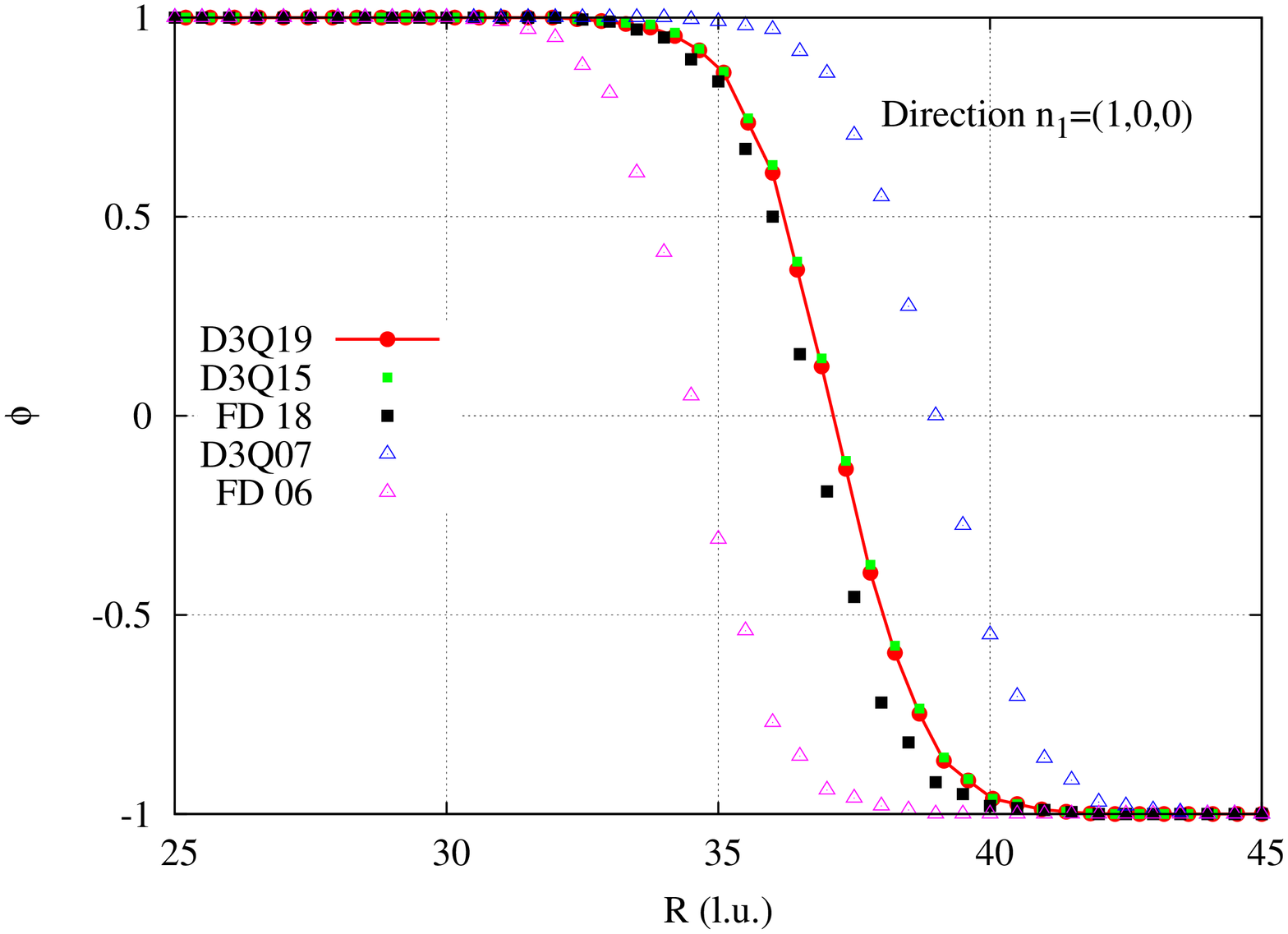} &  & \includegraphics[angle=-90,scale=0.33]{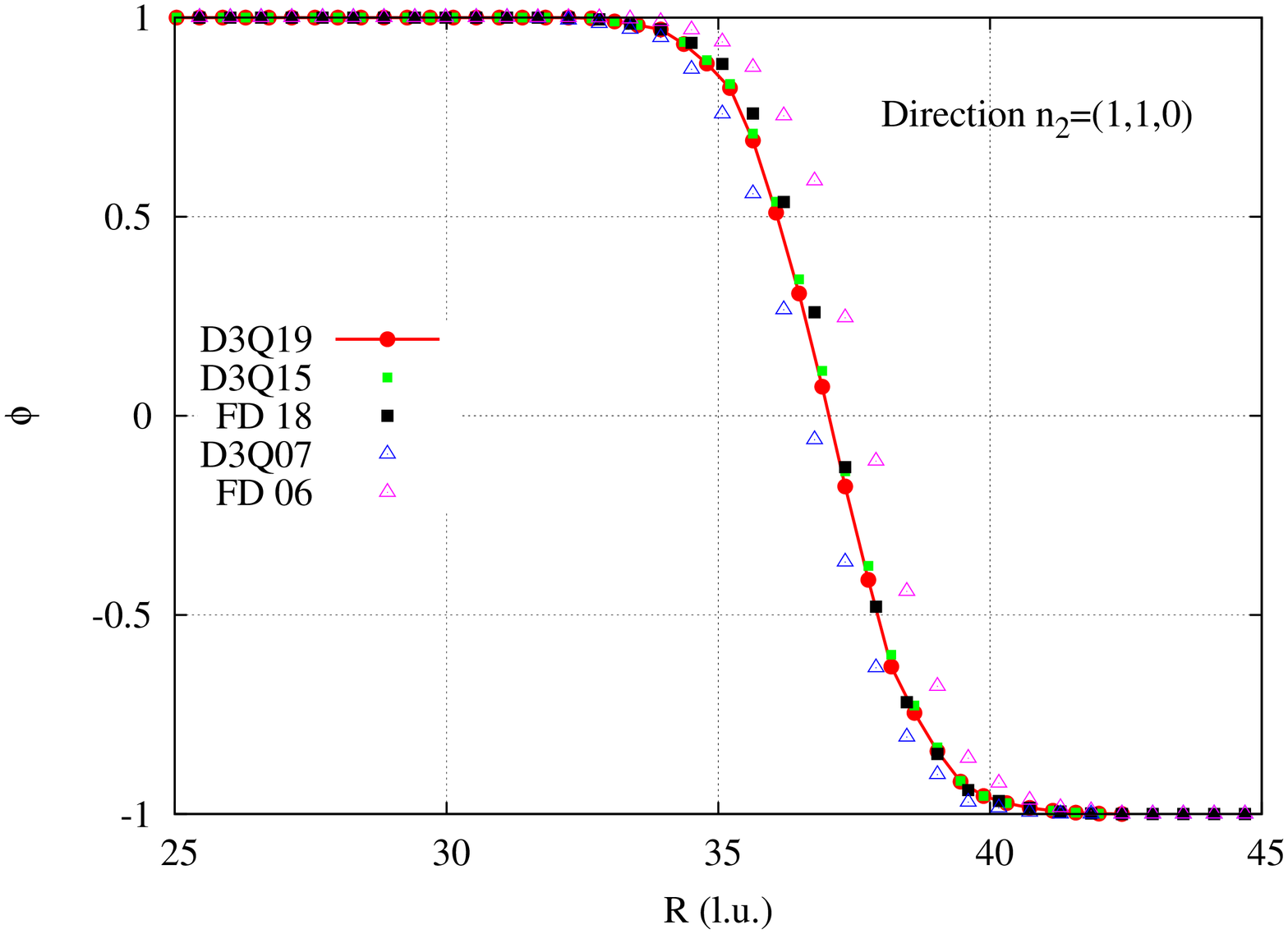}\tabularnewline
\end{tabular}
\par\end{centering}

~

~

\caption{\label{fig:Profils-dir_n1_n2}Phase-field profiles for directions
$\mathbf{n}_{1}$ (a) and $\mathbf{n}_{2}$ (b). Profiles calculated
by LB-D3Q15, LB-D3Q19 and FD18 fit for both directions unlike those
obtained by LB-D3Q7 and FD6.}
\end{figure*}

For this simulation, the mesh is composed of $301^{3}$ nodes, the
space-step is equal to $\delta x=0.01$ and the time-step is $\delta t=1.5\times10^{-5}$.
The interface thickness is equal to $W_{0}=0.01$, the scale factor
in time is $\tau_{0}=10^{-4}$. Finally $\varepsilon_{s}=0.05$, $\lambda=10$
and $\kappa=0.7$. The sphere radius is equal to $R_{s}=8$ lattice
unit (l.u.). The iso-values $\phi=0$ of the phase field are presented
in Fig. \ref{fig:Slices_D3Q7-D3Q19} at $t=10^{4}\delta t$ for three
lattices D3Q7, D3Q15 and D3Q19. In this figure, the initial condition
and the results obtained with FD6 and FD18 are plotted for comparison.
Slices are made for two different planes: the normal vector of the
first one is $(0,\,0,\,1)$ (Fig. \ref{fig:Slices_D3Q7-D3Q19}a),
and the normal vector of the second one is $(1,\,0,\,-1)$ (Fig. \ref{fig:Slices_D3Q7-D3Q19}b).
In each figure, the shapes of solutions obtained by LB-D3Q15, LB-D3Q19
and FD18 are circles that overlap, contrary to those obtained by LB-D3Q7
and FD6. The profiles collected along the directions $\mathbf{n}_{1}=(1,\,0,\,0)$
and $\mathbf{n}_{2}=(1,\,1,\,0)$ (Fig. \ref{fig:Profils-dir_n1_n2}),
present more accurately the effects of <<grid anisotropy>> (or mesh
anisotropy) of those two latter methods. The grid anisotropy can be
quantified by introducing a coefficient $\mathcal{A}_{g}$ \citep{Nestler_etal_JCP2005}:
$\mathcal{A}_{g}=\left|(R_{100}-R_{110})/(R_{100}+R_{110})\right|\times100$,
where $R_{100}$ is the radius measured along the $x$-axis in the
$\mathbf{n}_{1}$-direction and $R_{110}$ is the radius measured
at 45\textdegree{} between of the $x$-axis in the $\mathbf{n}_{2}$-direction.
The value of $\mathcal{A}_{g}$ is lower than one percent ($0.073$\%)
for D3Q15 and is equal to $0.116$\% for D3Q19. For D3Q7, the grid
anisotropy is equal to $3.095$\%. For the finite difference schemes,
$\mathcal{A}_{g}$ is equal to $4.365$\% for FD6 and $0.575$\% for
FD18. Results obtained with the D3Q15 lattice are slightly more accurate
because it is well-suited when the solidification occurs as a sphere.
Indeed, that lattice takes into account the diagonals of the cube
and allows the displacement of distribution function $f_{i}$ in the
diagonal directions, contrary to the D3Q19 (see Fig. 1b,c).

\paragraph{Anisotropic case: $\varepsilon_{s}\neq0$}

Now the validation of the numerical implementation is carried out
by considering an anisotropic case. We use for the comparison a 2D
numerical code based on a Finite Difference (FD) method for the phase-field
equation and a Monte-Carlo (MC) algorithm for the temperature \citep{Plapp-Karma_JCP2000}.
In what follows, the results of this method will be labeled by FDMC.
For the LB schemes, we use the lattices D2Q9 for Eq. (\ref{eq:PhaseField_KarmaRappel})
and D2Q5 for Eq. (\ref{eq:Temp_Ramirez}). The results will be labeled
by LBE.

The domain is a square discretized with meshes of size $\delta x$.
The initial seed is a diffuse circle of radius $R_{s}=10\delta x$
which is set at the origin of the computational domain. The problem
is symmetrical with respect to the $x$-axis and $y$-axis. In this
test, we compare the shape of the dendrite given by $\phi=0$ and
the evolution of the tip velocity. The interface thickness $W_{0}$
and the characteristic time $\tau_{0}$ are set to $W_{0}=\tau_{0}=1$.
The space step is chosen such as $\delta x/W_{0}=0.4$ \citep{Karma-Rappel_PRE1998},
the time step is $\delta t=0.008$ and the lengths of the system depend
on the undercooling $\Delta=-\theta_{0}$. A smaller undercooling
necessitates a bigger mesh because of the larger diffusive length.
The time to reach the stationary velocity is also more important.
We present below the results for two undercoolings: $\Delta_{1}=0.30$
and $\Delta_{2}=0.55$. For the first one, we use a mesh of $1000^{2}$
nodes and for the second one, a mesh of $500^{2}$ nodes.

In the phase-field theory, the capillary length $d_{0}$ and the kinetic
coefficient $\beta$ are given by \citep{Karma-Rappel_PRE1998}: $d_{0}=a_{1}W_{0}/\lambda$
and $\beta=a_{1}(\tau_{0}/\lambda W_{0}-a_{2}W_{0}/\kappa)$ where
$a_{1}=0.8839$ and $a_{2}=0.6267$. In this benchmark, we choose
the parameter $\lambda$ such as $\beta=0$, i.e. $\lambda^{\star}=\kappa\tau_{0}/a_{2}W_{0}^{2}$.
By considering $W_{0}=1$ and $\tau_{0}=1$, the coefficient $\lambda^{\star}$
is equal to $\lambda^{\star}=\kappa/a_{2}=1.59566\kappa$. For a thermal
diffusivity equals to $\kappa=4$, we obtain $\lambda^{\star}=6.3826$
and $d_{0}=0.1385$. Finally the anisotropic strength is $\varepsilon_{s}=0.05$.

\begin{figure*}
\begin{centering}
\begin{tabular}{ccc}
{\footnotesize (a) } &  & {\footnotesize (b)}\tabularnewline
\includegraphics[angle=-90,scale=0.35]{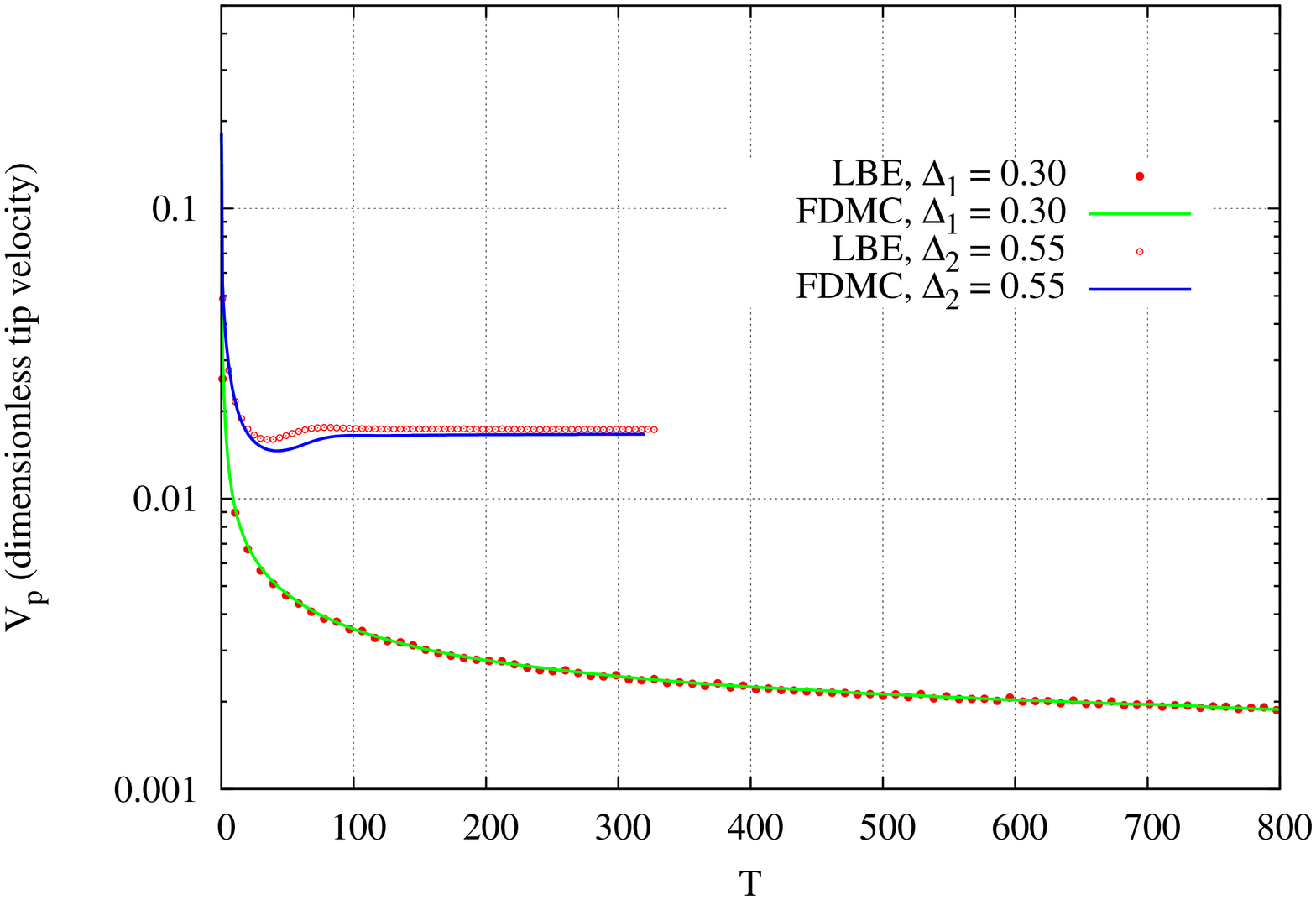} & $\quad$ & \includegraphics[angle=-90,scale=0.35]{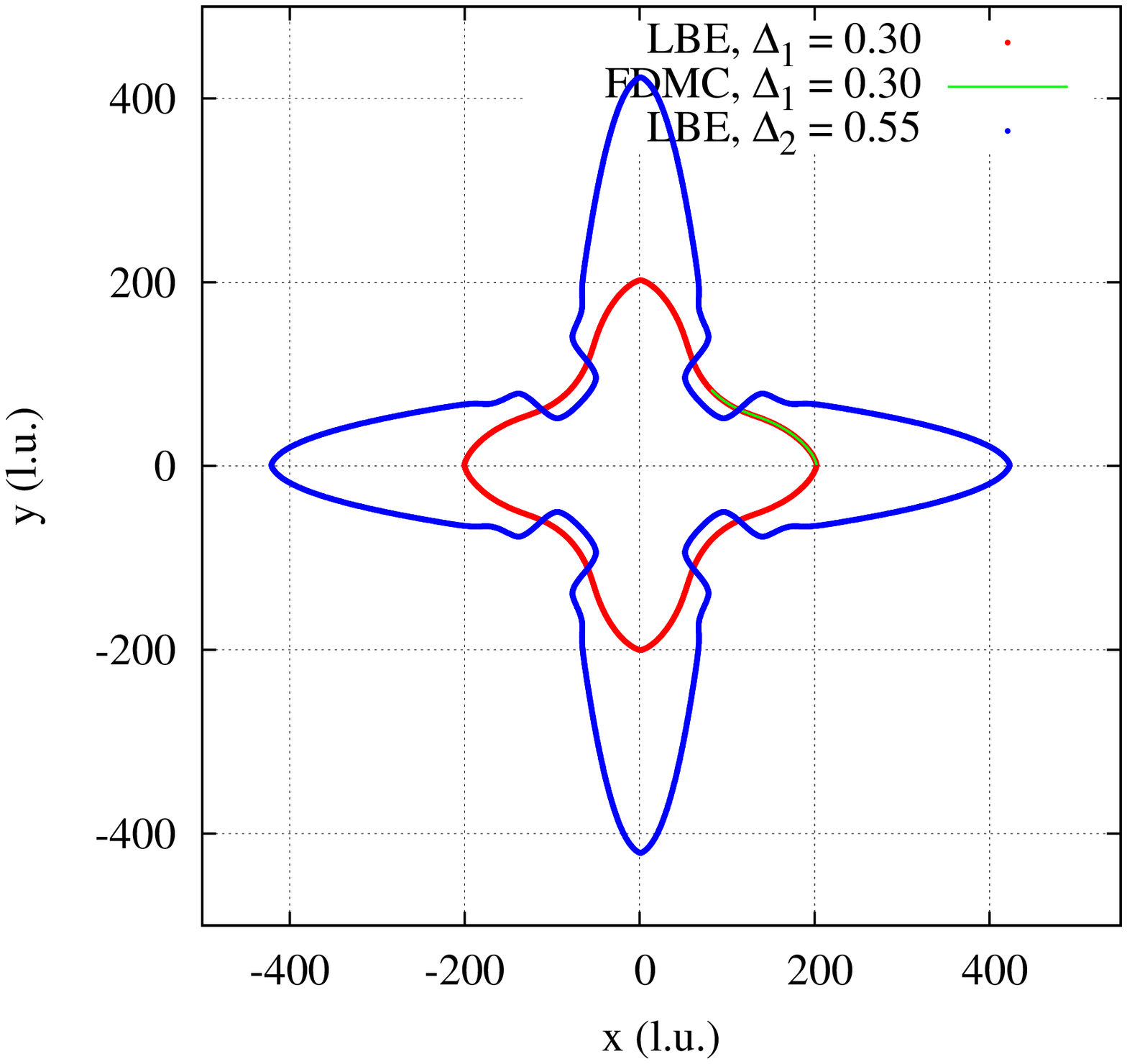}\tabularnewline
\end{tabular}
\par\end{centering}

~

~

\caption{\label{fig:Vitesse_Delta030}(a) Dimensionless tip velocity $V_{p}$
as a function of time for $\Delta_{1}=0.30$ and $\Delta_{2}=0.55$.
(b) Superposition of $\phi=0$ for FDMC (green line) and LBE (red
dots) at $t=1.3\times10^{5}\delta t$ for $\Delta_{1}$. For $\Delta_{2}$,
the shape $\phi=0$ is given for comparison at $t=4\times10^{4}\delta t$.
Parameters are $\kappa=4$, $\lambda^{\star}=6.3826$, $d_{0}=0.1385$
and $\varepsilon_{s}=0.05$.}
\end{figure*}

In the comparisons, the velocity $V_{p}$ is dimensionless by using
the factor $d_{0}/\kappa$ ($V_{p}=\tilde{V}_{p}d_{0}/\kappa$), the
position $x$ is also dimensionless by using the space-step ($x=\tilde{x}/\delta x$)
and the time $T$ is the time $t$ divided by $\tau_{0}$ ($T=t/\tau_{0}$).
Fig. \ref{fig:Vitesse_Delta030} presents the results of comparisons
for $\Delta_{1}=0.30$ and $\Delta_{2}=0.55$. For each numerical
method, the tip velocity fits well (Fig. \ref{fig:Vitesse_Delta030}a)
as well as the dendrite shape (Fig. \ref{fig:Vitesse_Delta030}b).
On this figure, the full dendrite is reconstructed by symmetry for
the LBE method. For the FDMC method, only the first quadrant is presented.
For $\Delta_{2}=0.55$, we remark a slight difference between both
curves during the initial transient that precedes steady-state growth
(in a time range from $t=0$ to $t=100$), but the steady-state velocities
converge toward values that are close to each other. Indeed, at $t=300$,
$V_{p}^{LBE}=0.01735$ and $V_{p}^{FDMC}=0.01667$, representing a
relative error of 4\%. For this benchmark, let us emphasize that the
value of $V_{p}$ reported in \citep{Karma-Rappel_PRE1998} (Table
II) is $V_{p}=0.0174$ and $V_{p}^{GF}=0.0170$ (where the $GF$ notation
stands for the Green's Function method, which is a sharp-interface
method considered as a reference), representing a relative error of
0.3\% between LBE and the first value, and 2\% between LBE and the
second value.

\subsection{Validation of supersaturation LB scheme}

\begin{figure*}
\begin{centering}
\includegraphics[angle=-90,scale=0.35]{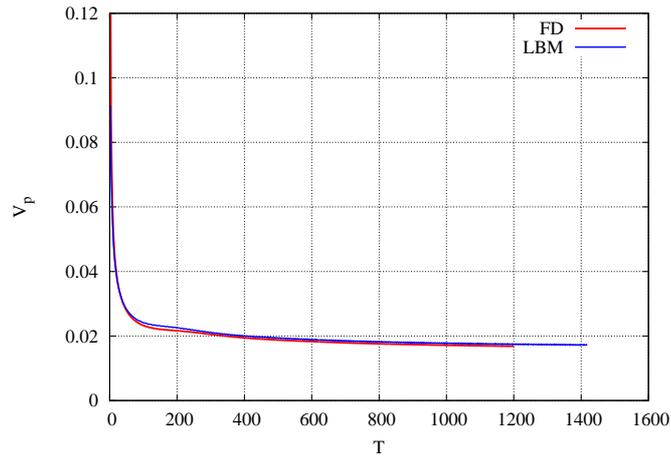}
\par\end{centering}

~

~

\caption{\label{fig:Vp_U0=00003D0.55}Dimensionless tip velocity $V_{p}$ of
an isothermal dilute alloy dendrite as a function of time for $U_{0}=-0.55$.}
\end{figure*}

The LB scheme for the supersaturation equation defines a new equilibrium
distribution function (Eq. (\ref{eq:Feq_Transport})) and necessitates
to calculate additional gradients in Eqs. (\ref{eq:Def_Jtot}) and
(\ref{eq:Def_S}). In order to check this method, an additional benchmark
is carried out by combining Eq. (\ref{eq:PhaseField_KarmaRappel})
coupled with Eq. (\ref{eq:Concentration_Echebarria}) including the
anti-trapping current $\mathbf{j}_{\mbox{at}}$ (Eq. (\ref{eq:AntiTrapping_Current})).
For this benchmark we consider an isothermal solidification, i.e.
$\theta=0$, and the parameters are $U_{0}=-0.55$, $D=2$, $k=0.15$,
$W_{0}=1$, $\tau_{0}=1$, $\varepsilon_{s}=0.03$, $\lambda^{\star}=3.2$,
$d_{0}=0.2762$, $R_{s}=10$ l.u., $Mc_{\infty}=1$, $\delta x=0.4$,
and $\delta t=0.02$.The LB results are compared with a finite-difference
code that is comparable to the one used in \citep{Karma_PRL2001}.
The tip velocity is presented in Fig. \ref{fig:Vp_U0=00003D0.55};
the good agreement validates the lattice Boltzmann scheme with anti-trapping
current.

\subsection{\label{sec:Simulations}Simulations of non standard dendrites}

The anisotropy function (Eq. (\ref{eq:Anisotropic-Function_As}))
defines an interfacial excess free energy which favors a preferential
growth in the direction {[}100{]}. Those directions correspond to
the directions of the main axes $x$, $y$ and $z$. Other preferential
directions of growth can be simulated by modifying this function on
the basis of spherical and cubic harmonics \citep{Karma_Orientation_Nature2006}.
In the present section, we compare the classical function (\ref{eq:Anisotropic-Function_As})
with another one defined by \citep{Hoyt-Asta-Karma_MatSciEng2003}:

\begin{equation}
a_{s}(\mathbf{n})=1+\varepsilon_{s}\left(\sum_{\alpha=x,y,z}n_{\alpha}^{4}-\frac{3}{5}\right)+\delta\left(3\sum_{\alpha=x,y,z}n_{\alpha}^{4}+66n_{x}^{2}n_{y}^{2}n_{z}^{2}-\frac{17}{7}\right).\label{eq:As_K41-K61}
\end{equation}

\begin{figure*}
\begin{centering}
\begin{tabular}{ccc}
{\footnotesize (a)} &  & {\footnotesize (b)}\tabularnewline
\includegraphics[scale=0.25]{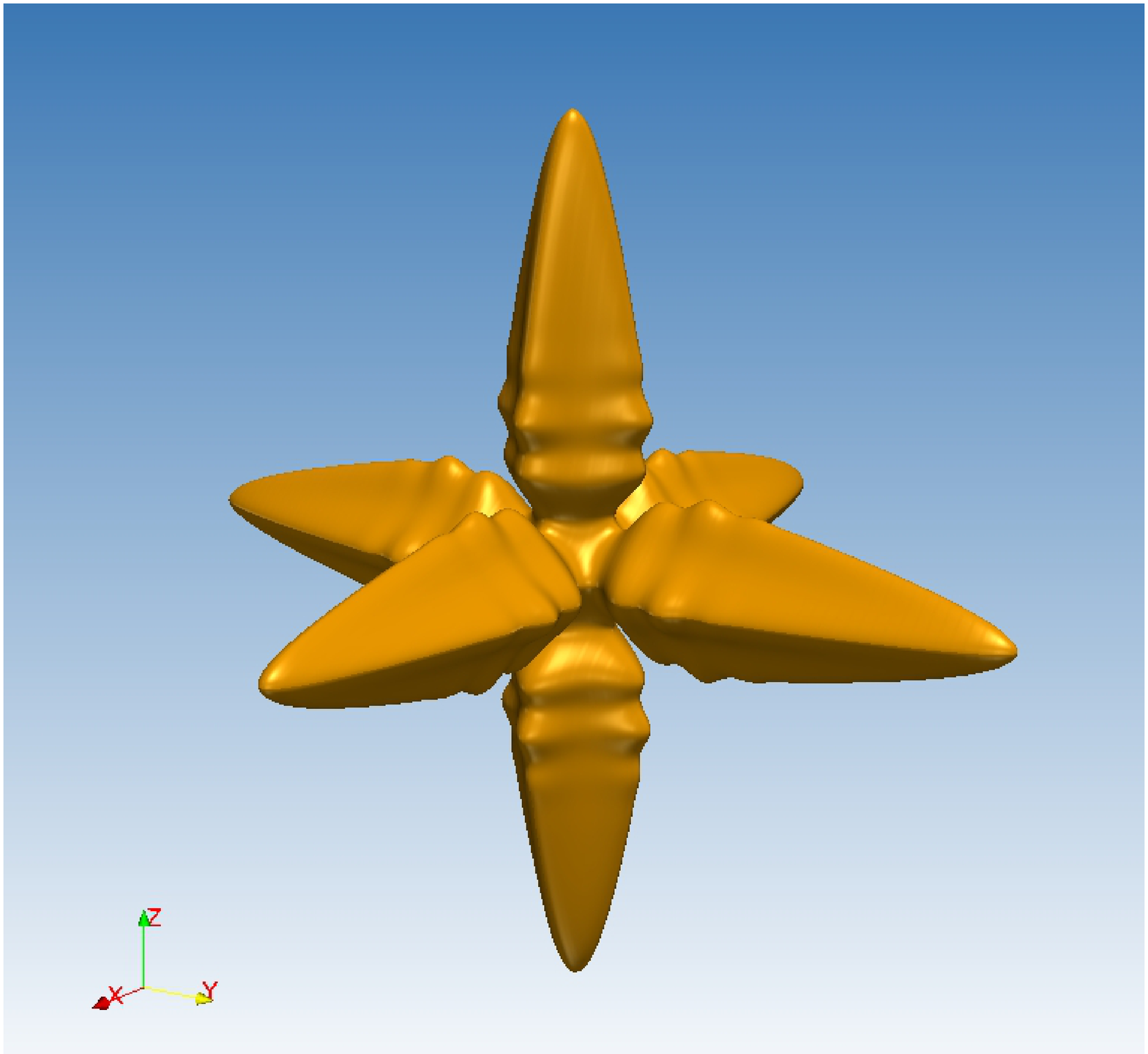} & $\qquad$ & \includegraphics[scale=0.25]{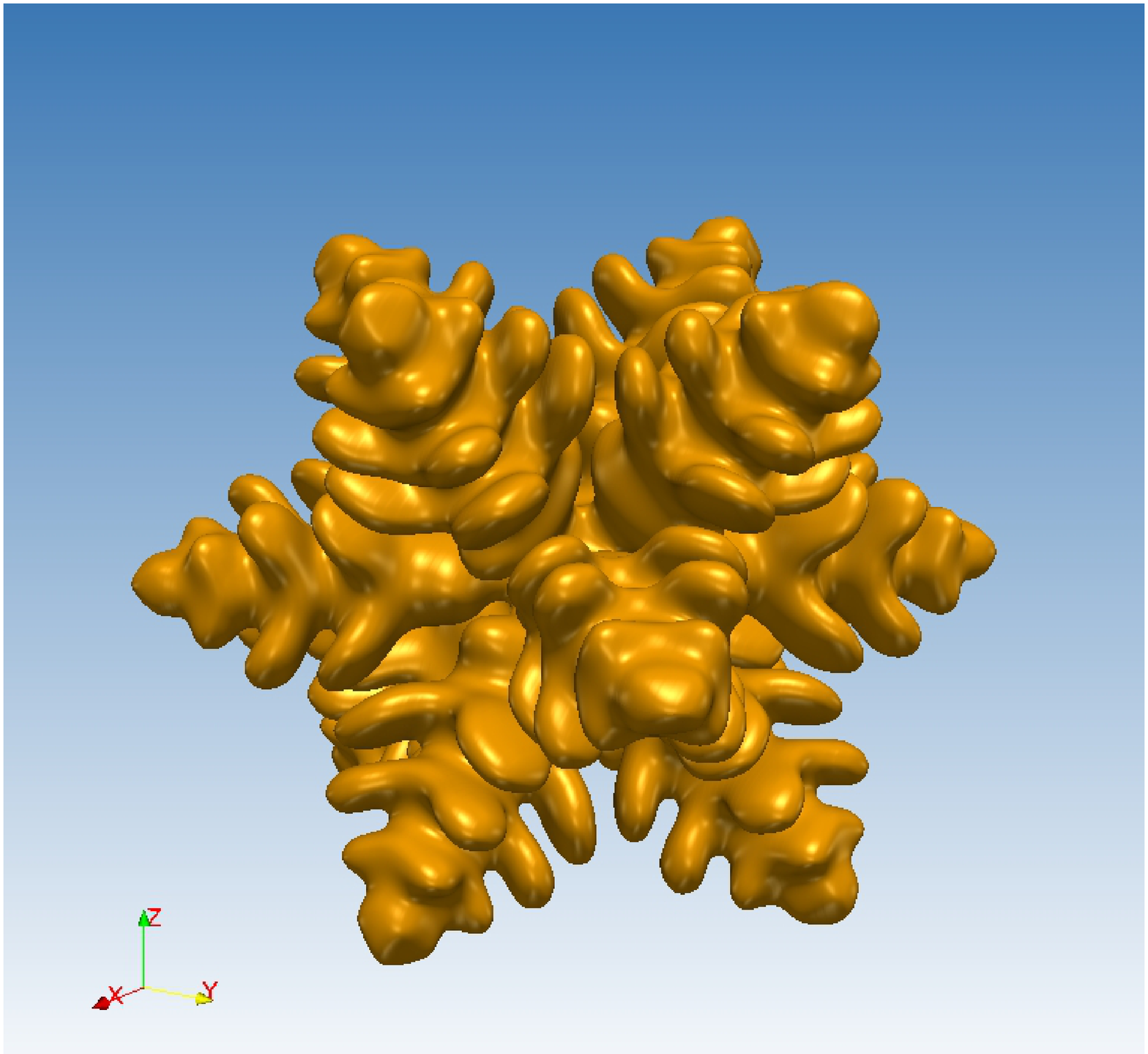}\tabularnewline
\end{tabular}
\par\end{centering}

\caption{\label{fig:Comparisons_As}(a) {[}100{]} preferential growth at $t=3\times10^{4}\delta t$
with $\varepsilon_{s}=0.05$. (b) {[}110{]} preferential growth at
$t=1.4\times10^{5}\delta t$ with $\varepsilon_{s}=0$ and $\delta=-0.02$.}
\end{figure*}

The second term in the right-hand side of Eq. (\ref{eq:As_K41-K61})
is the cubic harmonic $K_{41}$ and the last term corresponds to the
cubic harmonic $K_{61}$. In the LB method, the $a_{s}(\mathbf{n})$
function and its derivatives are involved in the function $\boldsymbol{\mathcal{N}}(\mathbf{x},\, t)$
inside the equilibrium distribution function Eq. (\ref{eq:Feq_Phase}).
For both simulations the kinetic coefficient is chosen such as $\tau(\mathbf{n})=\tau_{0}a_{s}^{2}(\mathbf{n})$,
the mesh is composed of $351^{3}$ nodes, $\delta x=0.01$, $\delta t=1.5\times10^{-5}$,
$W_{0}=0.0125$, $\tau_{0}=1.5625\times10^{-4}$, $\lambda=10$, $\kappa=1$
and $\Delta=0.25$. The first simulation is carried out by using Eq.
(\ref{eq:Anisotropic-Function_As}) and $\varepsilon_{s}=0.05$, and
the second one with Eq. (\ref{eq:As_K41-K61}), $\varepsilon_{s}=0$
and $\delta=-0.02$. The system is initialized with a sphere of radius
$R_{s}=8$ l.u. at the origin of the domain. The problem is symmetrical
with respect to the planes $xy$, $xz$ and $yz$. A comparison of
the shapes $\phi=0$ is presented in Fig. \ref{fig:Comparisons_As}
for a same orientation of the landmark. The method is thus able to
simulate easily different crystal shapes by modifying the function
$\boldsymbol{\mathcal{N}}(\mathbf{x},\, t)$.

\section{\label{sec:Conclusion}Conclusion}

We have presented a lattice Boltzmann method to simulate a crystal
growth model for a binary mixture with anti-trapping current. The
method requires a modification of the equilibrium distribution functions
and needs to consider a non-local collision for the phase-field equation
to take into account respectively the term responsible for the anisotropic
growth and the kinetic coefficient in front of the time derivative.
The use of lattices D3Q15 and D3Q19 for the phase-field equation improves
the accuracy of the solutions by removing the undesired effect of
grid anisotropy. The method was validated by comparison with other
codes based on the finite-difference method. Finally, the method is
able to simulate other anisotropic functions with minor modifications
of the code in order to generate preferential directions of dendritic
growth other than {[}100{]}.

The numerical method presented in this paper for the solidification
of alloys under diffusive heat and solute transport uses the same
concepts as those involved in the simulation of fluid flows: the lattices
(D2Q9, D3Q15, D3Q19) are identical and the same stages of collision,
displacement and bounce back are applied. This will make it easier
to directly couple the phase-field model and the Navier-Stokes equations
in order to study, for example, the density change effect during the
solidification process or the effect of convective fluid flow on crystal
growth. The advective terms that have to be added in each equation
of the phase-field model, can be taken into account by modifying the
equilibrium distribution functions of each equation according to standard
procedures. Studies including such couplings will be the subject of
future works.

\appendix

\section{\label{sec:Chapman-Enskog-PhaseField}Chapman-Enskog expansions for
phase-field equation}

We present in this appendix the Chapman-Enskog expansions for the
phase-field equation. In the first part, the continuous equation for
the moments of the equilibrium distribution function $g_{i}^{(0)}$
is established. In the second part, we focus on the derivation of
a specific form of the equilibrium distribution function $g_{i}^{(0)}$.
For more concision, dependencies in $\mathbf{x}$ and $t$ are canceled
in functions $g_{i}$, $\eta_{\phi}$ and $Q$. We also assume the
dependency of $a_{s}^{2}$ with $\mathbf{n}$.

\paragraph{\label{sub:Dvt-de-Taylor}Taylor and asymptotic expansions}

Taylor expansion at second-order in space and first-order in time
of Eq. (\ref{eq:LBE_Eq_Phase}) yields:

\begin{align}
a_{s}^{2}\left[g_{i}+\delta x\mathbf{e}_{i}\cdot\boldsymbol{\nabla}g_{i}+\frac{\delta x^{2}}{2}\mathbf{e}_{i}\mathbf{e}_{i}:\boldsymbol{\nabla}\boldsymbol{\nabla}g_{i}+\delta t\partial_{t}g_{i}\right] & =g_{i}+(a_{s}^{2}-1)\left[g_{i}+\delta x\mathbf{e}_{i}\cdot\boldsymbol{\nabla}g_{i}+\frac{\delta x^{2}}{2}\mathbf{e}_{i}\mathbf{e}_{i}:\boldsymbol{\nabla}\boldsymbol{\nabla}g_{i}\right]\nonumber \\
 & \qquad\qquad-\frac{1}{\eta_{\phi}}\left[g_{i}-g_{i}^{(0)}\right]+w_{i}Q_{\phi}\frac{\delta t}{\tau_{0}}\label{eq:Eq_Apres_Taylor}
\end{align}
After simplification, the factor $a_{s}^{2}$ appears only in front
of the time derivative $\partial_{t}g_{i}$:

\begin{align}
a_{s}^{2}\delta t\partial_{t}g_{i}+\delta x\mathbf{e}_{i}\cdot\boldsymbol{\nabla}g_{i}+ & \frac{\delta x^{2}}{2}\mathbf{e}_{i}\mathbf{e}_{i}:\boldsymbol{\nabla}\boldsymbol{\nabla}g_{i}=-\frac{1}{\eta_{\phi}}\left[g_{i}-g_{i}^{(0)}\right]+w_{i}Q_{\phi}\frac{\delta t}{\tau_{0}}.\label{eq:Expr_Apres_Taylor}
\end{align}
From now on all steps are standard (see \citep{Chen-Doolen_AnnRevFlMech1998,Walsh-Saar_WRR2010}).
Space and time are rescaled by introducing a small parameter $\epsilon=\delta x/L$
where $L$ is the characteristic length of the system. One scale in
space $\mathbf{x}_{1}=\epsilon\mathbf{x}$ is considered and two time-scales
$t_{1}=\epsilon t$ and $t_{2}=\epsilon^{2}t$ which are representative
of convection and diffusion, respectively. With these notations, the
partial derivatives write: $\boldsymbol{\nabla}=\epsilon\boldsymbol{\nabla}_{1}$
and $\partial_{t}=\epsilon\partial_{t_{1}}+\epsilon^{2}\partial_{t_{2}}$.
The function $g_{i}$ is expanded in power of $\epsilon$ around $g_{i}^{(0)}$:
$g_{i}\simeq g_{i}^{(0)}+\epsilon g_{i}^{(1)}.$ The moment of 0th-order
of the distribution function $g_{i}$ is the phase field $\phi$:
$\sum_{i}g_{i}=\phi$, which must be invariant during the collision
step. That means $\sum_{i}g_{i}^{(0)}=\phi$ and involves $\sum_{i}g_{i}^{(1)}=0$.
After substituting those relationships in (\ref{eq:Expr_Apres_Taylor}),
all terms in $\epsilon$- and those in $\epsilon^{2}$-order are combined
into two distinct equations. For the first one, the moment of zeroth-order
(sum over $i$) yields:

\begin{equation}
a_{s}^{2}\partial_{t_{1}}(\sum_{i}g_{i}^{(0)})+\frac{\delta x}{\delta t}\boldsymbol{\nabla}_{1}\cdot(\sum_{i}g_{i}^{(0)}\mathbf{e}_{i})=0,\label{eq:Eq_Ordre1_Final}
\end{equation}
and the moment of first-order (multiplying by $\mathbf{e}_{i}$ and
summing over $i$) yields:

\begin{equation}
\sum_{i}g_{i}^{(1)}\mathbf{e}_{i}\simeq-\eta_{\phi}\delta x\boldsymbol{\nabla}_{1}\cdot(\sum_{i}g_{i}^{(0)}\mathbf{e}_{i}\mathbf{e}_{i}).\label{eq:M1_g1}
\end{equation}
In (\ref{eq:M1_g1}), the term $\delta t\partial_{t_{1}}\sum_{i}g_{i}^{(0)}\mathbf{e}_{i}$
was assumed negligible, assumption that can be removed by modifying
the collision stage (see \citep{Zheng_etal_LargeDensityRatio_JCP2006}
for BGK-collision, \citep{ServanCamas-Tsai_AdWR2008} for TRT-collision
and \citep{dHumieres_etal_PhilTranRoySoc2002,Yoshida-Nagaoka_JCP2010}
for MRT-collision). For Eq. in $\epsilon^{2}$-order, by using (\ref{eq:M1_g1}),
the calculation of zeroth-order moment yields:

\begin{equation}
a_{s}^{2}\partial_{t_{2}}(\sum_{i}g_{i}^{(0)})=\boldsymbol{\nabla}_{1}\cdot\left[\left(\eta_{\phi}-\frac{1}{2}\right)\frac{\delta x^{2}}{\delta t}\boldsymbol{\nabla}_{1}\cdot(\sum_{i}g_{i}^{(0)}\mathbf{e}_{i}\mathbf{e}_{i})\right].\label{eq:Eq_Ordre2}
\end{equation}

Finally, by combining all terms $\epsilon^{0}\times\sum_{i}w_{i}Q_{\phi}/\tau_{0}$
+ $\epsilon^{1}\times$Eq. (\ref{eq:Eq_Ordre1_Final}) + $\epsilon^{2}\times$Eq.
(\ref{eq:Eq_Ordre2}), the continuous partial differential equation
for the three first moments of $g_{i}^{(0)}$ is:

\begin{align}
a_{s}^{2}\partial_{t}(\sum_{i}g_{i}^{(0)}) & =\boldsymbol{\nabla}\cdot\left[\left(\eta_{\phi}-\frac{1}{2}\right)\frac{\delta x^{2}}{\delta t}\boldsymbol{\nabla}\cdot(\sum_{i}g_{i}^{(0)}\mathbf{e}_{i}\mathbf{e}_{i})\right]-\frac{\delta x}{\delta t}\boldsymbol{\nabla}\cdot(\sum_{i}g_{i}^{(0)}\mathbf{e}_{i})+\sum_{i}w_{i}\frac{Q_{\phi}}{\tau_{0}}.\label{eq:Eq_g0_PhaseField}
\end{align}

\paragraph{\label{sub:Definition_G0_Eq}Equilibrium distribution function $g_{i}^{(0)}$}

Comparison of Eqs. (\ref{eq:Eq_g0_PhaseField}) and (\ref{eq:PhaseField_KarmaRappel})
that is rewritten as:

\begin{equation}
a_{s}^{2}\frac{\partial\phi}{\partial t}=\frac{W_{0}^{2}}{\tau_{0}}\boldsymbol{\nabla}\cdot(a_{s}^{2}(\mathbf{n})\boldsymbol{\nabla}\phi)+\frac{W_{0}^{2}}{\tau_{0}}\boldsymbol{\nabla}\cdot\boldsymbol{\mathcal{N}}+\frac{Q_{\phi}}{\tau_{0}},\label{eq:Eq_PhaseField_B}
\end{equation}
indicates that $g_{i}^{(0)}$ must be defined such that its moments
of 0th-, 1rst- and 2nd-order have to be equal to $\sum_{i}g_{i}^{(0)}=\phi$,
$\sum_{i}g_{i}^{(0)}\mathbf{e}_{i}=-\boldsymbol{\mathcal{N}}W_{0}^{2}\delta t/(\tau_{0}\delta x)$
and $\sum_{i}g_{i}^{(0)}\mathbf{e}_{i}\mathbf{e}_{i}=e^{2}\phi\overline{\overline{\mathbf{I}}}$
where $\overline{\overline{\mathbf{I}}}$ is the identity tensor of
rank 2. The equilibrium distribution function $g_{i}^{(0)}$ is chosen
as:

\begin{equation}
g_{i}^{(0)}=w_{i}\phi+w'_{i}\mathbf{e}_{i}\cdot\boldsymbol{\mathcal{N}}\frac{\delta t}{\delta x}\frac{W_{0}^{2}}{\tau_{0}},\label{eq:Hyp_g0}
\end{equation}
where we look for the coefficients $w_{i}$ and $w'_{i}$. Values
of weights and coefficient $e^{2}$ are detailed here for D3Q7 lattice
defined in section \ref{sec:Lattice-Boltzmann-method}. The generalization
for D3Q15 and D3Q19 lattices is straightforward. Moment of 0th-order
yields $\sum_{i}w_{i}\phi=\phi$ (the second term of the right-hand
side vanishes) and its moment of first-order yields:

\begin{equation}
\sum_{i}w_{i}\phi\mathbf{e}_{i}+\sum_{i}w'_{i}\left(\mathbf{e}_{i}\cdot\boldsymbol{\mathcal{N}}\frac{\delta t}{\delta x}\frac{W_{0}^{2}}{\tau_{0}}\right)\mathbf{e}_{i}=-\boldsymbol{\mathcal{N}}\frac{\delta t}{\delta x}\frac{W_{0}^{2}}{\tau_{0}},\label{eq:Contrainte_M1}
\end{equation}
where the first sum of the left-hand side vanishes. One obtains $w_{1}=w_{3}$,
$w_{2}=w_{4}$ and $w_{5}=w_{6}$. One solution is to set $w_{0}=1/4$
and $w_{i=1,\,...,\,6}=1/8$. Regarding the weights $w'_{k}$, the
following relationships are obtained by identifying the components
of each side of equality (\ref{eq:Contrainte_M1}): $(w'_{1}+w'_{3})\mathcal{N}_{x}=-\mathcal{N}_{x}$,
$(w'_{2}+w'_{4})\mathcal{N}_{y}=-\mathcal{N}_{y}$, $(w'_{5}+w'_{6})\mathcal{N}_{z}=-\mathcal{N}_{z}$.
We deduce that $w'_{i=1,\,...,\,6}=-1/2$. Calculation of second-order
moment of Eq. (\ref{eq:Hyp_g0}) yields (by using values of weights
$w_{i}$): $\sum_{i}g_{i}^{(0)}\mathbf{e}_{i}\mathbf{e}_{i}=\sum_{i}w_{i}\phi\mathbf{e}_{i}\mathbf{e}_{i}=(1/4)\phi\overline{\overline{\mathbf{I}}}.$
We set $e^{2}=1/4$ and $w'_{i}=w_{i}/e^{2}$, we obtain $g_{i}^{(0)}=w_{i}\left(\phi-e^{-2}\mathbf{e}_{i}\cdot\boldsymbol{\mathcal{N}}\delta tW_{0}^{2}/(\tau_{0}\delta x)\right)$.
Finally Eq. (\ref{eq:PhaseField_KarmaRappel}) is derived by identifying
$a_{s}^{2}W_{0}^{2}/\tau_{0}$ to $e^{2}(\zeta_{\phi}-1/2)\delta x^{2}/\delta t$.

\section{\label{sec:Appendix_Transport}Equilibrium distribution function
for the supersaturation equation}

Following the same procedure as detailed in \ref{sub:Dvt-de-Taylor},
the partial differential equation for the moments of $h_{i}^{(0)}$
is obtained:

\begin{align}
\partial_{t}(\sum_{i}h_{i}^{(0)}) & =\boldsymbol{\nabla}\cdot\left[\left(\eta_{U}-\frac{1}{2}\right)\frac{\delta x^{2}}{\delta t}\boldsymbol{\nabla}\cdot(\sum_{i}h_{i}^{(0)}\mathbf{e}_{i}\mathbf{e}_{i})\right]-\frac{\delta x}{\delta t}\boldsymbol{\nabla}\cdot(\sum_{i}h_{i}^{(0)}\mathbf{e}_{i})+\sum_{i}w_{i}\left[S+\frac{Q_{U}}{\zeta}\right].\label{eq:Eq_h0_PhaseField}
\end{align}

Comparison with Eq. (\ref{eq:Supersaturation_Bis}) indicates that
$h_{i}^{(0)}$ must be defined such as $\sum_{i}h_{i}^{(0)}=U$, $\sum_{i}h_{i}^{(0)}\mathbf{e}_{i}=\mathbf{J}_{\mbox{tot}}\delta t/\delta x$
and $\sum_{i}h_{i}^{(0)}\mathbf{e}_{i}\mathbf{e}_{i}=e^{2}(q(\phi)/\zeta(\phi))U\overline{\overline{\mathbf{I}}}$.
The equilibrium distribution function $h_{i}^{(0)}$ is set as follows:

\begin{equation}
h_{i}^{(0)}=A_{i}U+B_{i}\frac{q(\phi)}{\zeta(\phi)}U+C_{i}\mathbf{e}_{i}\cdot\mathbf{J}_{\mbox{tot}}\frac{\delta t}{\delta x},\label{eq:Hyp_h0}
\end{equation}
where coefficients $A_{i}$, $B_{i}$ and $C_{i}$ have to be determined.
Moment of 0th-order yields the first constraint $\sum_{i}A_{i}+B_{i}Dq(\phi)/\zeta(\phi)=1$
and the moment of 1rst-order yields the second one:

\begin{align}
\sum_{i}\left(A_{i}+B_{i}\frac{q(\phi)}{\zeta(\phi)}\right)U\mathbf{e}_{i}+\sum_{i}\left(C_{i}\mathbf{e}_{i}\cdot\mathbf{J}_{\mbox{tot}}\frac{\delta t}{\delta x}\right)\mathbf{e}_{i} & =\mathbf{J}_{\mbox{tot}}\frac{\delta t}{\delta x}.\label{eq:Constraint_M1}
\end{align}

One solution satisfying both equalities is: $A_{0}=1$, $A_{1,...,6}=0$,
$B_{0}=-3/4$, $B_{1,...,6}=1/8$ and $C_{1,...,6}=1/2$. Using values
of $A_{i}$ and $B_{i}$ for calculation of the 2nd-order moment,
we check that: $\sum_{i}h_{i}^{(0)}\mathbf{e}_{i}\mathbf{e}_{i}=(1/4)(q(\phi)/\zeta(\phi))U\overline{\overline{\mathbf{I}}}$.
We set $e^{2}=1/4$ and $C_{i}/B_{i}=1/e^{2}$. The expected supersaturation
Eq. (\ref{eq:Concentration_Echebarria}) is obtained by identifying
$D=e^{2}(\eta_{U}-1/2)(\delta x^{2}/\delta t)$.

\section*{Acknowledgements}

A. Cartalade wishes to thank the SIVIT project, involving AREVA, for
the financial support.

\bibliographystyle{elsarticle-num-names}
\bibliography{/media/Elements/REDAC/Biblio-BibTeX/Biblio-PhaseField,/media/Elements/REDAC/Biblio-BibTeX/Biblio-LBM,/media/Elements/REDAC/Biblio-BibTeX/Biblio-Solidification}

\end{document}